\documentclass[prd,twocolumn,superscriptaddress,floatfix,amsmath,amssymb,amsfonts,nofootinbib,longbibliography]{revtex4-2}

\usepackage{tcolorbox}
\usepackage{float} 
\usepackage{scalerel}
\usepackage[normalem]{ulem}
\usepackage[english]{babel}
\usepackage{graphicx}
\usepackage{dcolumn}

\usepackage{bm}
\usepackage{blindtext}
\usepackage{verbatim}
\usepackage{relsize}
\usepackage{mathrsfs}
\usepackage{musicography}
\usepackage{amsmath}
\usepackage{blindtext}
\usepackage{cancel}
\usepackage{physics}
\usepackage{epstopdf}
\usepackage{mathtools}
\usepackage{blindtext}
\usepackage{tensor}
\usepackage{color}
\usepackage[usenames,dvipsnames]{pstricks}
\usepackage{epsfig}
\usepackage{pst-grad} 
\usepackage{pst-plot} 
\usepackage{hyperref}
\usepackage{verbatim}
\usepackage{slashed}
\usepackage{dsfont}
\usepackage[english]{babel}
\usepackage{amsmath,amssymb,amsthm}  
\usepackage{leftindex,enumerate}
\usepackage{enumitem, kantlipsum}  
\usepackage{tikz}

\newcommand{\proj}[2]{\left| {#1} \right\rangle\!\left\langle {#2} \right|}

\newcommand{\mf}{\mathsf}

\newcommand{\ii}{\mathrm{i}}

\newcommand{\tc}[1]{\textsc{#1}}

\definecolor{goldenrod}{rgb}{0.85, 0.65, 0.13}

\allowdisplaybreaks[1] 

\begin{document}

\title{Nonlinear particle detectors across the Rindler firewall}

\author{Matheus H. Zambianco}
\email{mhzambia@uwaterloo.ca}

\affiliation{Department of Applied Mathematics, University of Waterloo, Waterloo, Ontario, N2L 3G1, Canada}
\affiliation{Institute for Quantum Computing, University of Waterloo, Waterloo, Ontario, N2L 3G1, Canada}
\affiliation{Perimeter Institute for Theoretical Physics, Waterloo, Ontario, N2L 2Y5, Canada}



\author{Eduardo Mart\'{i}n-Mart\'{i}nez}
\email{emartinmartinez@uwaterloo.ca}

\affiliation{Department of Applied Mathematics, University of Waterloo, Waterloo, Ontario, N2L 3G1, Canada}
\affiliation{Institute for Quantum Computing, University of Waterloo, Waterloo, Ontario, N2L 3G1, Canada}
\affiliation{Perimeter Institute for Theoretical Physics, Waterloo, Ontario, N2L 2Y5, Canada}

\begin{abstract}
We investigate Unruh-DeWitt detectors coupled to composite observables of a quantum scalar field, including quadratic coupling to the field momentum and coupling to the local energy density. We develop a distributional framework for evaluating the corresponding detector response functions and apply it to detectors crossing the Rindler firewall. While we recover the finite response of the derivative-coupling model, we show that quadratic momentum coupling leads to products of distributions that do not admit a canonical, regulator-independent definition in the sharp-firewall idealization. The vacuum correspondence between the local energy-density and quadratic momentum responses motivates the conjecture that analogous difficulties may arise for energy-density coupling in the sharp firewall model. Within the sharp, pointlike composite-coupling model considered here, our analysis further suggests that these pathologies originate from the discontinuous severing of correlations across the Rindler horizon.
\end{abstract}

\maketitle

\section{Introduction}

Particle detector models provide a versatile operational framework for implementing localized measurements and quantum information protocols in quantum field theory (QFT). One of their main advantages is that they describe localized interactions between quantum fields and finite-dimensional quantum systems, thereby avoiding many of the spurious divergences that arise when attempting to define localized field observables directly. The simplest, yet remarkably powerful example is the Unruh-DeWitt (UDW) detector~\cite{Unruh1976,DeWitt,Unruh-Wald,Schlicht,Jorma}, which (in its simplest version) consists of a two-level quantum system coupled to a quantum field within a finite spacetime region. UDW-like models have been extensively used in relativistic quantum information and in foundational studies in QFT, including entanglement harvesting in flat~\cite{Valentini1991, Reznik2003, Pozas-Kerstjens:2015, Pozas2016, morotebalboa2026} and curved~\cite{Nick,HarvestingBHLaura, KeithRobEdu2018, ericksonBH, ampEntBH2020, hectorMass, Membrere_2023, wang2025harvestinginformationhorizon, ZambiancoAdamEdu2026, wurtz2026} spacetimes, quantum energy teleportation~\cite{teleportation, hotta2011intro, QETExperiment}, quantum collect calling~\cite{Jonsson_2015, collectCalling, PRLHyugens2015, PetarEduCosmological} to name a few. They have also proved useful in establishing and clarifying central theoretical phenomena such as the Unruh effect~\cite{Unruh1976,Takagi,matsasUnruh,LoukoUnruh} and Hawking radiation~\cite{Unruh1976,HawkingBHbook,JormaHawking}.

One of the simplest measurable quantities associated with a particle detector is its response function, which gives the probability that the detector becomes excited through its interaction with the field \cite{birrell_davies, Schlicht}. The dependence of this quantity on the quantum state of the field has been investigated in a variety of settings, including the vacuum \cite{Schlicht, Jorma}, thermal \cite{thermalresponse, Costa_1995}, and coherent states \cite{cohrentresponseold, coherentharvestingPetarEdu}.

In this work, following the approach of \cite{Louko2014}, we analyze a less standard situation in which the field is prepared in a state whose correlations across the Rindler horizon mimic the properties of the firewall proposal~\cite{firewall_main, PhysRevLett.110.101301}, which is motivated by the black hole information loss problem: if black hole evaporation is to preserve unitarity, then the smooth near-horizon vacuum of Hawking's original description~\cite{HawkingBHbook} might have to be replaced by a region of violent structure, a ``firewall'', that would destroy infalling observers while severing correlations between field modes on opposite sides of the horizon~\cite{firewall_main, Mathur_2009, PhysRevLett.110.101301}.

The resulting \emph{firewall state} is locally indistinguishable from the Minkowski vacuum within each Rindler wedge, but the correlations across the Rindler horizon are removed. In \cite{Louko2014}, it was shown that a single UDW detector linearly coupled to a massless scalar field (or its time derivative) prepared in such a state experiences a finite response when crossing the Rindler horizon. Subsequently, it was also shown that the entanglement between two initially correlated UDW detectors is not destroyed when one of them crosses the Rindler firewall~\cite{EduJormaPRLFire}. These results suggest that, despite its singular character, the firewall state may not be as violent to matter crossing it and that it can be probed in a controlled way by standard detector models.

A natural question, then, is whether this good behavior persists for matter coupling to arbitrary field observables, for example, when one considers interaction models more directly tied to non-linear observables of physical interest, such as the field’s energy density. This question is particularly relevant because the firewall state is not Hadamard~\cite{Louko2014,fullingHadamard,fewsterNecessityHadamard}---presenting divergences in the stress-energy tensor across the firewall --- and one may therefore expect new singular features to appear once the detector couples to composite operators rather than to the field itself. Detector models coupled to stress-energy observables have been studied before (see, e.g., \cite{strees_energy_coupling_1987}), where pointlike accelerated detectors coupled to components of the stress-energy tensor were analyzed, and explicit excitation rates were derived for the trace coupling. Here, we consider a pointlike detector coupled to the local energy density of the field as measured in its instantaneous rest frame.

To do so, we first show that, for inertial pointlike detectors in the Minkowski vacuum, the energy-density response is proportional to the quadratic momentum response, motivating the latter as a simpler model for the present analysis. Using this simpler model, we study the detector's response when the field is prepared in the firewall state. First, we re-derive the results of \cite{Louko2014}, treating the two-point functions explicitly as distributions in position space, where the discontinuity of the firewall two-point function is manifest. We then find that, in the sharp, pointlike quadratic-momentum model, the formal response involves products of distributions that have no canonical, regulator-independent definition. These products arise explicitly from derivatives of the discontinuous wedge-indicator functions, identifying the sharp firewall as a source of the obstruction in this calculation. This mechanism suggests that smoothing the firewall may remove this particular source of ill-definedness.

This paper is organized as follows. In Sec.~\ref{sec:Minkowski_vacuum}, we introduce the energy-density coupling model and show that, for the purposes of the present analysis, its essential features are captured by a model in which the detector couples to the square of the field momentum. In Sec.~\ref{sec:firewall}, we revisit the results of Ref.~\cite{Louko2014} using an approach based on distributional methods, focusing on the derivative-coupling (momentum) interaction model. Our main results are presented in Sec.~\ref{sec:andthenitdiverges}, where we show that the response function in the quadratic momentum model exhibits persistent pathologies. Finally, we summarize our conclusions in Sec.~\ref{sec:conclusions}.

\section{The essential features of the energy density coupling}
\label{sec:Minkowski_vacuum}

We consider a two-level particle detector with an energy gap $\Omega$ coupled to the stress-energy density of a scalar, massless field in $n$+1-dimensional Minkowski spacetime. Let $\mathsf{u} = \partial_{\tau}$ denote the four-velocity associated to the detector centre-of-mass motion for a Fermi-Walker rigid detector (see, for instance, \cite{EduRickBrunosmearedgeneral, us2}), and $\hat{T}_{\mu \nu}$ the energy-momentum tensor of the field. Then, the interaction Hamiltonian density that dictates the time evolution of the system (in the interaction picture) is
\begin{equation}
    \hat{h}_{\tc{i}}(\mf x) = \lambda \Lambda(\mf x)\hat{\mu}(\tau) u^{\mu}u^{\nu}\hat{T}_{\mu \nu}(\mf x),\label{eq:energy_density_coupling}
\end{equation}
where $\Lambda(\mf x)$ is the spacetime smearing function, $\lambda$ is the coupling strength, and the detector's monopole moment is given by
\begin{equation}
    \hat{\mu}(\tau) = e^{\ii \Omega \tau} |e \rangle  \langle g| + e^{-\ii \Omega \tau}|g \rangle \langle e|,
\end{equation}
where $\tau$ is the proper time associated with the detector's center of mass and $\ket{g}$, $\ket{e}$ are the ground and excited states, respectively. With this convention, allowing $\Omega<0$ amounts to interchanging the labels of the two energy eigenstates.
We recall that for a massless scalar field $\hat{\phi}$ in Minkowski spacetime, the energy-momentum tensor reads
\begin{equation}
    \hat{T}_{\mu \nu} = \partial_{\mu}\hat{\phi}\partial_{\nu}\hat{\phi} - \frac{1}{2}\eta_{\mu \nu}\partial^{\beta}\hat{\phi}\partial_{\beta}\hat{\phi}
\end{equation}
We choose inertial coordinates $\mf x = (t, \bm x) \equiv (t, x^{i})$ and assume that the detector moves inertially $\mathsf{u} = \partial_{t}$. In this case, the Hamiltonian density becomes
\begin{equation}
  \hat{h}_{\tc{i}}(\mf x)= \frac{\lambda}{2}\Lambda(\mf x)\hat{\mu}(t)((\partial_{t}\hat{\phi})^2 + \partial_{i}\hat{\phi}\partial^{i}\hat{\phi}),
\end{equation}
To compute time evolution, let us assume that at some initial time $t_0$ sufficiently in the past, the detector-field system is prepared in the following state:
\begin{equation}
    \hat{\rho}_{0} = \proj{g}{g}
 \otimes \proj{0}{0},
 \label{eq:initial_state_global}
\end{equation}
where  $\ket{0}$ is the field's Minkowski vacuum. The unitary implementing time evolution is then given by
\begin{equation}
\hat{U}_{\tc{i}} = \mathcal{T} \exp \left(-\ii \int{\dd v \, \hat{h}_{\tc{i}}(\mf x)} \right),
\end{equation}
where $\dd v \,$ is the invariant spacetime volume element, which in the coordinates we chose reads $\dd v \, = \dd t \,\dd^n  \bm{x}$. After the interaction, the reduced state of the detectors reads
\begin{equation}
    \hat{\rho}_{\tc{d}} = \text{Tr}_{\phi}[\hat{U}_{\tc{i}}\hat{\rho}_{0}\hat{U}^{\dagger}_{\tc{i}}] = \proj{g}{g} + \hat{\rho}^{(1)}_{\tc{d}} + \hat{\rho}^{(2)}_{\tc{d}} + \mathcal{O}(\lambda^3),
\end{equation}
where we performed a perturbative expansion in the parameter $\lambda$. To control divergences, we make the assumption that in the vacuum, the expected value of the stress energy density should be zero (compatible with saying that in the vacuum state, $\ket{0}$, we observe no spacetime curvature). This translates in the usual prescription of normal ordering for the renormalized stress-energy density:
\begin{equation}
      : \hat{T}_{\mu\nu}(\mf x) :  \,\equiv\, \hat{T}_{\mu\nu}(\mf x)- \bra{0} \hat{T}_{\mu\nu}(\mf x)  \ket{0}\openone.
    \label{eq:mean_value_zero}
\end{equation}
At leading order in the coupling strength, the state of the detector density matrix after the interaction can be represented in the basis $\{\ket{g}, \ket{e}\}$ as
\begin{equation}
    \hat{\rho}^{(\tc{e})}_{\tc{d}} =  \begin{bmatrix}
{ 1 - \mathcal{L}^{(\tc{e})}_{0}} & 0 \\
0 & {\mathcal{L}^{(\tc{e})}_{0}}  \\ 
\end{bmatrix}
+  {\cal O}(\lambda ^{4}),
\label{rho_D}
\end{equation}
where the superindex $(\text{E})$ notates the coupling to the energy density. The {\it excitation probability} reads
\begin{align}
\mathcal{L}^{(\tc{e})}_{0} = & \lambda^2 \int\dd v \, \dd v \,' \Lambda(\mf x)\Lambda(\mf x')e^{-\ii \Omega (t - t')} \times \nonumber \\ & \langle 0 | \!:\!\hat{T}_{tt}(\mf x)\!: :\! \hat{T}_{tt}(\mf x')\!:\!| 0 \rangle \nonumber \\ & \equiv \lambda^2 \mathcal{F}_{0}^{(\tc{e})},
\label{eq:L_term_vanilla}
\end{align}
where $\mathcal{F}_{0}^{(\tc{e})}$ is the so-called {\it response function}. It turns out that the energy density correlation function, \mbox{$\langle 0 | \!:\!\hat{T}_{tt}(\mf x)\!: :\! \hat{T}_{tt}(\mf x')\!:\!| 0 \rangle$} can be written in terms of the vacuum Wightman function,
\begin{equation}
    W_{0}(\mf x, \mf x') = \langle 0 | \hat{\phi}(\mf x)\hat{\phi}(\mf x') | 0 \rangle.
\end{equation}
Indeed, one can show that (see Appendix \ref{sec:appendix_two_point})
\begin{align}
\label{eq:stress_energy_two_point_general}
    & \langle 0|\!:\!\hat{T}_{tt}(\mf x)\!: :\!\hat{T}_{tt}(\mf x')\!:\!|0\rangle  \\*\nonumber
    &= \frac{1}{2}\Bigl[
        \bigl(\partial_{t}\partial_{t'}W_{0}(\mf x,\mf x')\bigr)^2
        + \sum_{i}\bigl(\partial_{i}\partial_{t'}W_{0}(\mf x,\mf x')\bigr)^2 \\*
    &\quad\;{}
        + \sum_{i}\bigl(\partial_{t}\partial_{i'}W_{0}(\mf x,\mf x')\bigr)^2
        + \sum_{i,j}\bigl(\partial_{i}\partial_{j'}W_{0}(\mf x,\mf x')\bigr)^2
    \Bigr].\nonumber    
\end{align}
Looking at Eq.~\eqref{eq:stress_energy_two_point_general}, one might wonder whether one can find a simpler interaction model which essentially displays the same behaviour as that of $\mathcal{F}^{(\tc{e})}$. In particular, let us analyze the quadratic momentum coupling, where the detector couples to the (renormalized) square of the canonically conjugate momentum of the field amplitude $\hat{\pi}(\mf x) = \partial_{\tau}\hat{\phi}(\mf x)$. Namely, we consider the Hamiltonian density:
\begin{equation}
    \hat{h}_{\tc{i}}(\mf x) = \lambda \Lambda(\mf x)\hat{\mu}(\tau)\mathopen:\hat{\pi}(\mf x)^2 \mathopen{:}, 
    \label{eq:physical_interaction}
\end{equation}
where the renormalized squared momentum is
\begin{equation}
    \mathopen{:} \hat{\pi}^2(\mf x)\mathopen{:} \equiv \hat{\pi}^2(\mf x) - \bra{0} \hat{\pi}^2(\mf x)  \ket{0}\openone.
\label{eq:normal_ordering_standard}
\end{equation}
We assume the same setup as in the energy density coupling, with the initial state of the system detector-field given by Eq.~\eqref{eq:initial_state_global}. We also fix inertial coordinates $\mf x = (t, \boldsymbol{x})$, where the detector's proper time coincides with the global inertial time, $t = \tau$. In this setup, the response function, now denoted by $\mathcal{F}_{0}^{(\pi^2)}$, requires the evaluation of the correlation function \cite{eduAchimBosonFermion,Sachs1} (See Appendix~\ref{sec:appendix_two_point}).
\begin{equation}
    \langle 0| \mathopen:\hat{\pi}(\mf x)^2 \mathopen{:} \mathopen{:}\hat{\pi}(\mf x')^2\mathopen{:} | 0 \rangle = 2 (\partial_{t}\partial_{t'}W_{0}(\mf x, \mf x'))^2.
\label{eq:correlation_momentum_squared}
\end{equation}
Thus,
\begin{equation}
    \mathcal{F}^{(\pi^2)}_{0} = 2\int\dd v \, \dd v \,' \Lambda(\mf x)\Lambda(\mf x')e^{-\ii \Omega (t - t')} (\partial_{t}\partial_{t'}W_{0}(\mf x, \mf x'))^2.
\end{equation}
Next, we shall study and compare the behavior of the response functions $\mathcal{F}^{(\tc{e})}_{0}$ and $\mathcal{F}^{(\pi^2)}_{0}$ considering pointlike detectors in $1 + 1$ and $ 3 + 1$ dimensions.

\subsection{1+1 Minkowski}
\label{sec:Mink1_1}
For $ 1 + 1$-dimensional Minkowski spacetime, the stress-energy density two-point function, Eq. \eqref{eq:stress_energy_two_point_general}, reduces to
\begin{align}
    & \langle 0 | \!:\!\hat{T}_{tt}(\mf x)\!: :\! \hat{T}_{tt}(\mf x')\!:\!| 0 \rangle = \frac{1}{2} [(\partial_{t}\partial_{t'}W_{0}(\mf x, \mf x'))^2 \nonumber \\ & + (\partial_{t}\partial_{x'}W_{0}(\mf x, \mf x'))^2
     + (\partial_{x}\partial_{t'}W_{0}(\mf x, \mf x'))^2 \nonumber \\ & + (\partial_{x}\partial_{x'}W_{0}(\mf x, \mf x'))^2)],
\label{eq:two_point_func_tmunu}
\end{align}
Because we are working with a scalar, massless quantum field in 1 + 1 Minkowski, the Wightman function requires an IR cutoff $\Lambda$~\cite{birrell_davies}. The expression for the Wightman function in this case is \cite{birrell_davies}
\begin{equation}
    W_{0}(\mf x, 
    \mf x') =-\frac{1}{4 \pi} \log \left(-\Lambda^2\left[(\Delta t-\mathrm{i} \epsilon)^2-\Delta x^2\right]\right),
    \label{eq:Wightman_IR}
\end{equation}
with $\Delta x = x - x'$ and $\Delta t = t - t'$. Recall that all the observables involving $W(\mf x, \mf x')$ should be evaluated in the limit $\epsilon \to 0^+$, where the Wightman function is properly understood as a distribution. Thus, throughout this paper, unless we explicitly evaluate an observable (e.g., the detector's response function), we shall keep the term $\ii \epsilon$ as a convenient notation.

\begin{figure}
    \centering
    \includegraphics[width=1.0\linewidth]{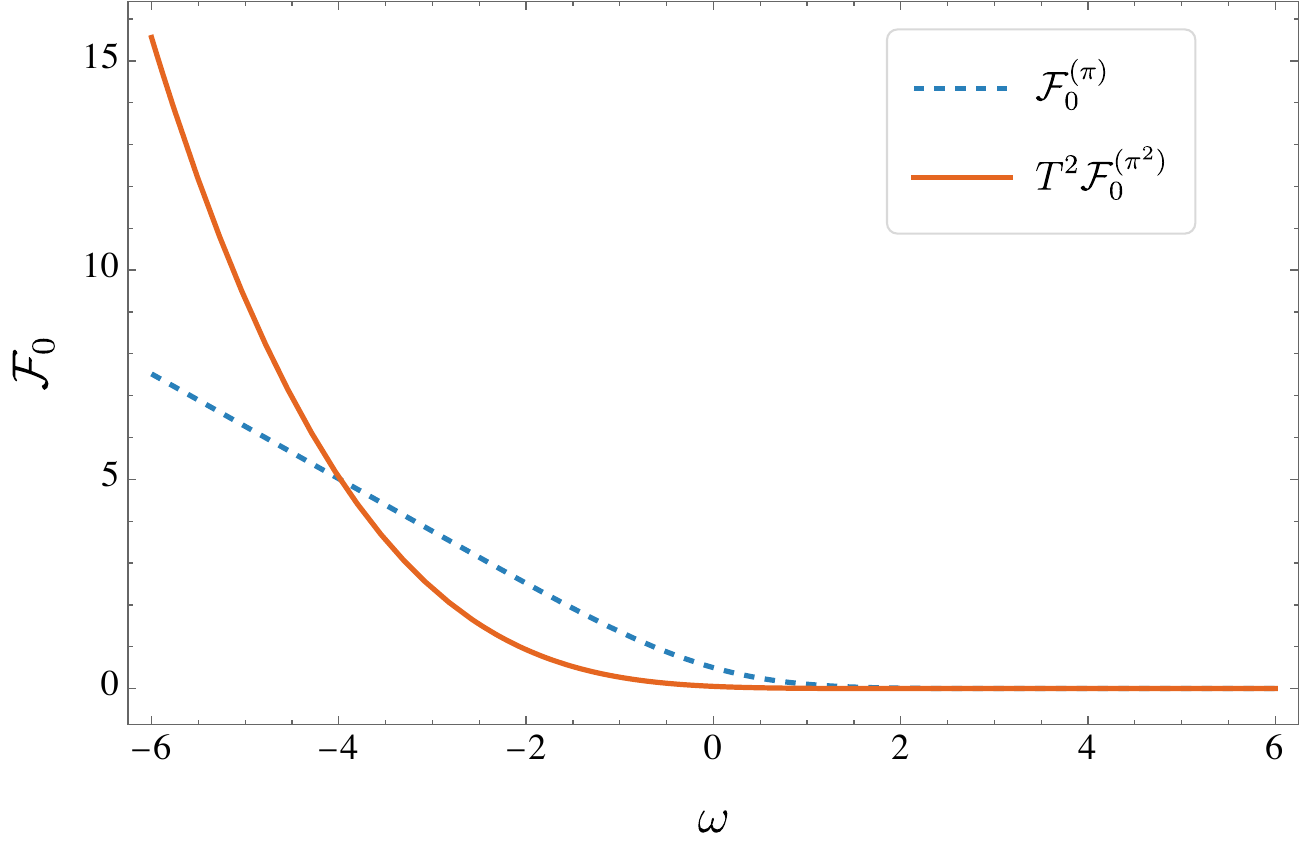}
   \caption{Comparison between the (dimensionless) response functions $\mathcal{F}^{(\pi)}_{0}$ and $T^2\mathcal{F}^{(\pi^2)}_{0}$ for Gaussian switching, shown as functions of $\omega=\Omega T$ for $1 + 1$-dimensional Minkowski}
    \label{fig:example1_plus_1}
\end{figure}

As expected, the derivatives in Eq. \eqref{eq:two_point_func_tmunu} eliminate the dependence on the IR cutoff, so there is no arbitrariness in the predictions of the theory. Indeed, using Eq.~\eqref{eq:stress_energy_two_point_general} one can evaluate
\begin{align}
    & \langle 0 | \!:\!\hat{T}_{tt}(\mf x)\!: :\! \hat{T}_{tt}(\mf x')\!:\!| 0 \rangle  = \nonumber \\
    & = \frac{1}{8 \pi^2}\left[\frac{1}{(\Delta t - \Delta x - \ii \epsilon)^4} + \frac{1}{(\Delta t + \Delta x - \ii \epsilon)^4}\right]. 
    \label{eq:two_point_vanilla_tmunu}
\end{align}
Similarly, one finds
\begin{equation}
     \langle 0| \mathopen{:}\hat{\pi}(\mf x)^2 \mathopen{:} \mathopen{:}\hat{\pi}(\mf x')^2\mathopen{:} | 0 \rangle =
\frac{(\Delta x^{2}+(\Delta t - \ii \epsilon)^{2})^2}
{2\pi^{2}(\Delta x^{2}-(\Delta t - \ii \epsilon)^{2})^{4}}.
\label{eq:two_point_vanilla_pi2}
\end{equation}
Next, we consider a pointlike detector with a generic, smooth switching $\chi(t)$, following an inertial worldline $\mf z(t)  = (t, x_{0})$,
\begin{equation}
    \Lambda(t, x) = \chi(t)\delta(x - x_{0}).
    \label{eq:pointlike_smearing}
\end{equation}
In this case, the pull-back of the correlation functions \eqref{eq:two_point_vanilla_tmunu} and \eqref{eq:two_point_vanilla_pi2} to the detector's trajectory reads
\begin{equation}
    \langle 0 | \!:\!\hat{T}_{tt}(t, x_{0})\!: :\! \hat{T}_{tt}(t', x_{0})\!:\!| 0 \rangle = \frac{1}{4\pi^2}\frac{1}{(\Delta t - \ii \epsilon)^4}, 
\end{equation}
and
\begin{equation}
    \langle 0| \mathopen{:}\hat{\pi}(t, x_{0})^2 \mathopen{:} \mathopen{:}\hat{\pi}(t', x_{0})^2\mathopen{:} | 0 \rangle = \frac{1}{2\pi^2} \frac{1}{(\Delta t - \ii \epsilon)^4}.
\end{equation}
Therefore, in this setup the response functions $\mathcal{F}^{(\tc{e})}_{0}$ and $\mathcal{F}^{(\pi^2)}_{0}$ are basically the same (except for a factor of 2 that could be reabsorbed in the coupling strength for the actual probabilities), that is
\begin{equation}
    \mathcal{F}^{(\tc{e})}_{0} = \frac{1}{2}\mathcal{F}^{(\pi^2)}_{0}.
\end{equation}
Let us now obtain  a closed expression for $\mathcal{F}^{(\pi^2)}_{0}$. We have
\begin{equation}
    \mathcal{F}^{(\pi^2)}_{0} = \frac{1}{2\pi^2} \lim_{\epsilon \to 0^{+}}\int_{-\infty}^{\infty}\dd t \, \int_{-\infty}^{\infty}\dd t \,' \chi(t) \chi(t') \frac{e^{-\ii \Omega \Delta t}}{(\Delta t - \ii \epsilon)^4},
\end{equation}
where the limit $\epsilon \to 0^{+}$ is understood in the distributional sense.
Performing the change of variables $(t, t') \to (u, s)$, with $u = t - t'$ and $s = t$, we obtain
\begin{equation}
     \mathcal{F}^{(\pi^2)}_{0} = \lim_{\epsilon \to 0^{+}}\frac{1}{2\pi^2} \int_{-\infty}^{\infty}\dd u \, \frac{\Phi(u)}{(u - \ii \epsilon)^4},
\end{equation}
where we defined the auxiliary function
\begin{equation}
    \Phi(u) \equiv e^{-\ii \Omega u}\int_{-\infty}^{\infty}\dd s \chi(s)\chi(s - u),
\end{equation}
which is simply the convolution product of the switching function, modulated by a complex phase that depends on the detector's energy gap $\Omega$. In the distributional manipulations below, we assume that the switching function $\chi$ is at least of class $C^{n-1}(\mathbb{R})$, and either has compact support or satisfies $\chi^{(k)}(t)\to0$ as $|t|\to\infty$ for $0\leq k\leq n-1$, with sufficient decay to ensure convergence of the relevant finite-part and principal-value integrals. For any $n \in \mathbb{N}$ it is possible to show that (see details in Appendix \ref{sec:Vanilla_calculations})
\begin{equation}
    \lim_{\epsilon \to 0}\frac{1}{(x \mp \ii \epsilon)^n} = \operatorname{FP}\left(\frac{1}{x^n} \right) \pm \frac{(-1)^{n - 1}}{(n - 1)!} \ii \pi \delta^{(n - 1)}(x),
    \label{eq:generalizedSP_main_text}
\end{equation}
where $\operatorname{FP}$ denotes the Hadamard Finite Part distribution, which can be defined by its relationship with the Principal Value distribution ($\operatorname{PV}$) when acting on a test function $f$:
\begin{equation}
    \left \langle \operatorname{FP}\left(\frac{1}{x^{n}}\right), f  \right \rangle = \frac{1}{(n - 1)!}\left \langle \operatorname{PV}\left(\frac{1}{x}\right), f^{(n-1)}  \right \rangle. 
\label{eq:FP_and_PV_main_text}
\end{equation}
The relationship of this identity with the more formal definition of Hadamard finite part can be found in Appendix \ref{sec:Vanilla_calculations}. For what follows, we recall that  the derivative of the Dirac delta acting on any test function $f$ is defined as
\begin{equation}
    \int_{-\infty}^{\infty}\dd x \, f(x)\delta^{(n)}(x) = (-1)^{n}f^{(n)}(0).
\end{equation}
Therefore, for any $n \in \mathbb{N}$ and any function $f$ of class $\mathcal{C}^{n - 1}(\mathbb{R})$, we have
\begin{equation}
    \lim_{\epsilon \to 0^{+}}\int_{-\infty}^{\infty}\dd x \, \frac{f(x)}{(x -\ii \epsilon)^n} = \frac{\ii \pi f^{(n - 1)}(0)}{(n - 1)!}  + \operatorname{FP}\int_{-\infty}^{\infty}\dd x \,\frac{f(x)}{x^n}.
    \label{eq:that_is_the_one}
\end{equation}
Thus, in our case ($n = 4$) we can write
\begin{equation}
    \mathcal{F}^{(\pi^2)}_{0} =  \frac{1}{12 \pi^2} \left[\ii \pi \Phi^{(3)}(0) + \operatorname{PV} \int_{-\infty}^{\infty} \dd u \, \frac{\Phi^{(3)}(u)}{u} \right].
    \label{eq:general_main_text}
\end{equation}
Under the standard physical assumption that the field-detector coupling smoothly vanishes in the far past and in the far future, 
\begin{equation}
    \lim_{t \to \pm \infty}\chi(t) = 0,
\end{equation}
we have
\begin{equation}
    \Phi^{(3)}(0) = \ii \Omega^3 \int_{-\infty}^{\infty}\dd s \, \chi(s)^2 + 3 \ii \Omega \int_{-\infty}^{\infty}\dd s \, \chi'(s)^2.
\end{equation}
As for the $\operatorname{PV}$ integral, we can write, for any function of $f$ class $\mathcal{C}^{1}(\mathbb{R})$,
\cite{GelfandShilov2016}
\begin{align}
    \operatorname{PV} \int_{-\infty}^{\infty}\dd x \, \frac{f(x)}{x} & = \lim_{\delta \to 0^{+}} \int_{\delta}^{\infty}\dd x \, \frac{f(x) - f(-x)}{x} \nonumber \\ & \equiv\int_{0}^{\infty}\dd x \, \frac{f(x) - f(-x)}{x}, 
\end{align}
where the last integral is understood as an improper integral. Thus, 
\begin{align}
    \mathcal{F}_{0}^{(\pi^2)} & =- \frac{\Omega^3}{12\pi}\int^{\infty}_{-\infty}\dd s \, \chi(s)^2 - \frac{\Omega}{4 \pi} \int^{\infty}_{-\infty}\dd s \, \chi'(s)^2 \nonumber \\ & + \frac{1}{12\pi^2}\int_{0}^{\infty}\dd u \, \frac{\Phi^{(3)}(u) - \Phi^{(3)}(-u)}{u}.
    \label{eq:pi2_mink_response_1_plus_1}
\end{align}
To write this fully in terms of the switching function and the detector's energy gap, let
\begin{equation}
    F(u) = \int_{-\infty}^{\infty}\dd s \chi(s)\chi(s - u).
\end{equation}
Clearly, $F(-u) = F(u)$, so $F$ is an even function. A direct calculation then yields
\begin{align}
\Phi^{(3)}(u)-\Phi^{(3)}(-u)
& =
2\cos(\Omega u)\bigl(F^{(3)}(u)-3\Omega^2 F'(u)\bigr)
\nonumber \\ & +
2\sin(\Omega u)\bigl(\Omega^3 F(u)-3\Omega F''
(u)\bigr).
\end{align}
To analyze a concrete example, we choose a Gaussian switching,
\begin{equation}
    \chi(t) = e^{-t^2/T^2}.
    \label{eq:gaussian_switching}
\end{equation}
We also compare the response function $\mathcal{F}^{(\pi^2)}_{0}$ with the response function of the derivative-coupling model, where the interaction Hamiltonian density is
\begin{equation}\label{eq:derivativeHamo}
    \hat{h}_{\tc{i}}(\mf x) = \lambda \Lambda(\mf x)\hat{\mu}(\tau)\hat{\pi}(\mf x),
\end{equation}
and the response function is given by (see, e.g., \cite{derivativeJorma}) 
\begin{equation}
 \mathcal{F}^{(\pi)}_{0}  = \int_{-\infty}^{\infty}\dd t \, \int_{-\infty}^{\infty}\dd t \,' \chi(t)\chi(t')\partial_{t}\partial_{t'}W_{0}(t, t').
 \label{eq:response_derivative_coupling_Mink}
\end{equation}
This response function can also be reduced to a one-variable integral, and then, evaluated distributionally using \eqref{eq:generalizedSP_main_text}. Indeed
\begin{align}
    \mathcal{F}^{(\pi)}_{0} & = -\frac{1}{2\pi}\int_{-\infty}^{\infty}\dd u \, \frac{\Phi(u)}{(u - \ii \epsilon)^2} \nonumber \\ & = -\frac{1}{2\pi}\Biggl[ \ii \pi \Phi'(0) + \operatorname{PV}\int_{-\infty}^{\infty}\dd u \, \frac{\Phi'(u)}{u} \Biggr] \nonumber \\ & = -\frac{\Omega}{2}\int_{-\infty}^{\infty}\dd s \chi(s)^2 \nonumber \\ & - \frac{1}{2\pi}\int_{0}^{\infty}\dd u \, \frac{\Phi'(u) - \Phi'(-u)}{u}.
\end{align}
Using the switching function of Eq.~\eqref{eq:gaussian_switching}, we obtain closed-form expressions for both response functions (with $\omega = \Omega T$):
\begin{equation}
    \mathcal{F}^{(\pi)}_{0} = \frac{1}{4}\left[2e^{-\frac{\omega^2}{2}}-\sqrt{2\pi}\,\omega \,\operatorname{erfc}\!\left(\frac{\omega}{\sqrt{2}}\right)\right],
\end{equation}
\begin{equation}
    \mathcal{F}^{(\pi^2)}_{0} = \frac{1}{T^2}\frac{2e^{-\frac{\omega^2}{2}}\left(2+\omega^2\right)-\sqrt{2\pi}\,\omega\left(3+\omega^2\right)\operatorname{erfc}\!\left(\frac{\omega}{\sqrt{2}}\right)}{24\pi}
\end{equation}
where $\operatorname{erfc}$ is the complementary error function, $\operatorname{erfc}(x) = 1 - \erf{x}$. In Fig. \ref{fig:example1_plus_1}, we compare the behavior of $\mathcal{F}^{(\pi)}_{0}$ and $\mathcal{F}^{(\pi^2)}_{0}$ as a function of $\omega = \Omega T$. Both response functions decay for large energy gaps, but they exhibit quantitatively different behavior near $\Omega=0$. For small $|\Omega|$ the response of detectors  coupled to $\hat{\pi}$ is generally larger. By contrast, for sufficiently negative\footnote{Recall $\Omega<0$ is equivalent to considering that the detector starts in the excited state instead of the ground state.} $\Omega$ the $\hat{\pi}^2$-coupled detectors have the larger response, thus having a larger propensity for de-excitation.

\subsection{3 + 1 Minkowski}

In a 3 + 1-dimensional Minkowski spacetime with coordinates $\mf x = (t, \bm x)$, the vacuum Wightman function reads
\begin{equation}
    W_{0}(\mf x, \mf x') = \frac{1}{4\pi^2(-(\Delta t - \ii \epsilon)^2 + |\Delta \bm x|^2)},
\end{equation}
with $\Delta t = t - t'$, $\Delta \bm x = \bm x - \bm x'$. The relevant correlation functions are given by
\begin{align}
   &  \langle 0 | \!:\!\hat{T}_{tt}(\mf x)\!: :\! \hat{T}_{tt}(\mf x')\!:\!| 0 \rangle =  \nonumber \\ & \frac{
    \left(|\Delta \boldsymbol{x}|^2 + 3(\Delta t-\ii\epsilon)^2\right)
    \left(3|\Delta \boldsymbol{x}|^2 + (\Delta t-\ii\epsilon)^2\right)
    }{
    2\pi^4\left(|\Delta \boldsymbol{x}|^2-(\Delta t-\ii\epsilon)^2\right)^6
    }.
\end{align}
and
\begin{equation}
    \langle 0 | \!:\!\hat{\pi}^2(\mf x)\!: :\! \hat{\pi}^2(\mf x')\!:\!| 0 \rangle =
\frac{\left(|\Delta \boldsymbol{x}|^{2}+3\left(\Delta t-\ii\epsilon\right)^{2}\right)^{2}}{2\pi^{4}\left(|\Delta \boldsymbol{x}|^{2}-\left(\Delta t - \ii\epsilon\right)^{2}\right)^{6}}.
\end{equation}
Once more, we consider a pointlike detector following the inertial trajectory $\mf z(t) = (t, \bm x_{0})$. In this case, pulling back the correlation functions to the detector's trajectory yields
\begin{equation}
    \langle 0 | \!:\!\hat{T}_{tt}(t, \boldsymbol{x}_{0})\!: :\! \hat{T}_{tt}(t', \boldsymbol{x}_{0})\!:\!| 0 \rangle = \frac{3}{2\pi^4} \frac{1}{ (\Delta t - \ii \epsilon)^8},
\end{equation}
and
\begin{equation}
    \langle 0 | \!:\!\hat{\pi}^2(t, \boldsymbol{x}_{0})\!: :\! \hat{\pi}^2(t', \boldsymbol{x}_{0})\!:\!| 0 \rangle = \frac{9}{2 \pi^4} \frac{1}{(\Delta t - \ii \epsilon)^8}.
\end{equation}
From this, one immediately sees that
\begin{equation}\label{eq:propto_responses}
\mathcal{F}^{(\tc{e})}_{0} = \frac{1}{3}\mathcal{F}^{(\pi^2)}_{0}. 
\end{equation}
Therefore, in this setting, the physics of the energy density coupling is essentially the same as that of the quadratic momentum coupling. It is not hard to see that a proportionality relationship $\mathcal{F}^{(\tc{e})}_{0} \propto\mathcal{F}^{(\pi^2)}_{0}$ will hold in any spacetime dimensions. In $n + 1$ Minkowski spacetime with $n > 1$, the vacuum Wightman (massless) is (see, e.g, \cite{Takagi, ericksonNew})
\begin{equation}
    W_{0}(\mf x, \mf x') =
    \frac{(-\ii)^{n-1}\Gamma\!\left(\frac{n-1}{2}\right)}
    {4\pi^{\frac{n+1}{2}}\left[(\Delta t-\ii\epsilon)^2-|\Delta \boldsymbol{x}|^2\right]^{\frac{n-1}{2}}}.
\end{equation}
Then, in the pointlike case, both correlation functions \eqref{eq:stress_energy_two_point_general} and \eqref{eq:correlation_momentum_squared} will scale as
\begin{equation}
    \frac{1}{(\Delta t - \ii \epsilon)^{2n + 2}}.
\end{equation}
The evaluation of the response function $\mathcal{F}^{(\pi^2)}_{0}$ follows exactly the same technique applied in Sec.~\ref{sec:Mink1_1}. Indeed, one can show that
\begin{equation}
\mathcal{F}_{0}^{(\pi^2)} = \frac{9}{2\pi^4}\int_{-\infty}^{\infty}\dd u \, \frac{\Phi(u)}{(u - \ii \epsilon)^8}.
\end{equation}
Then, using identities \eqref{eq:generalizedSP_main_text} and \eqref{eq:FP_and_PV_main_text} with $n = 8$, we obtain
\begin{equation}
\mathcal{F}_{0}^{(\pi^2)} =   \frac{1}{1120 \pi^4}\Biggl[\ii \pi \Phi^{(7)}(0) + \operatorname{PV} \int_{-\infty}^{\infty}\dd u \, \frac{\Phi^{(7)}(u)}{u} \Biggr].
\label{eq:response_derivative_67}
\end{equation}
Due to the presence of the derivative of order 7, writing this explicitly in terms of the energy gap and the switching function (or the corresponding convolution product) is too cumbersome. We shall then leave the result in this form.

\begin{figure}
    \centering
\includegraphics[width=1.0\linewidth]{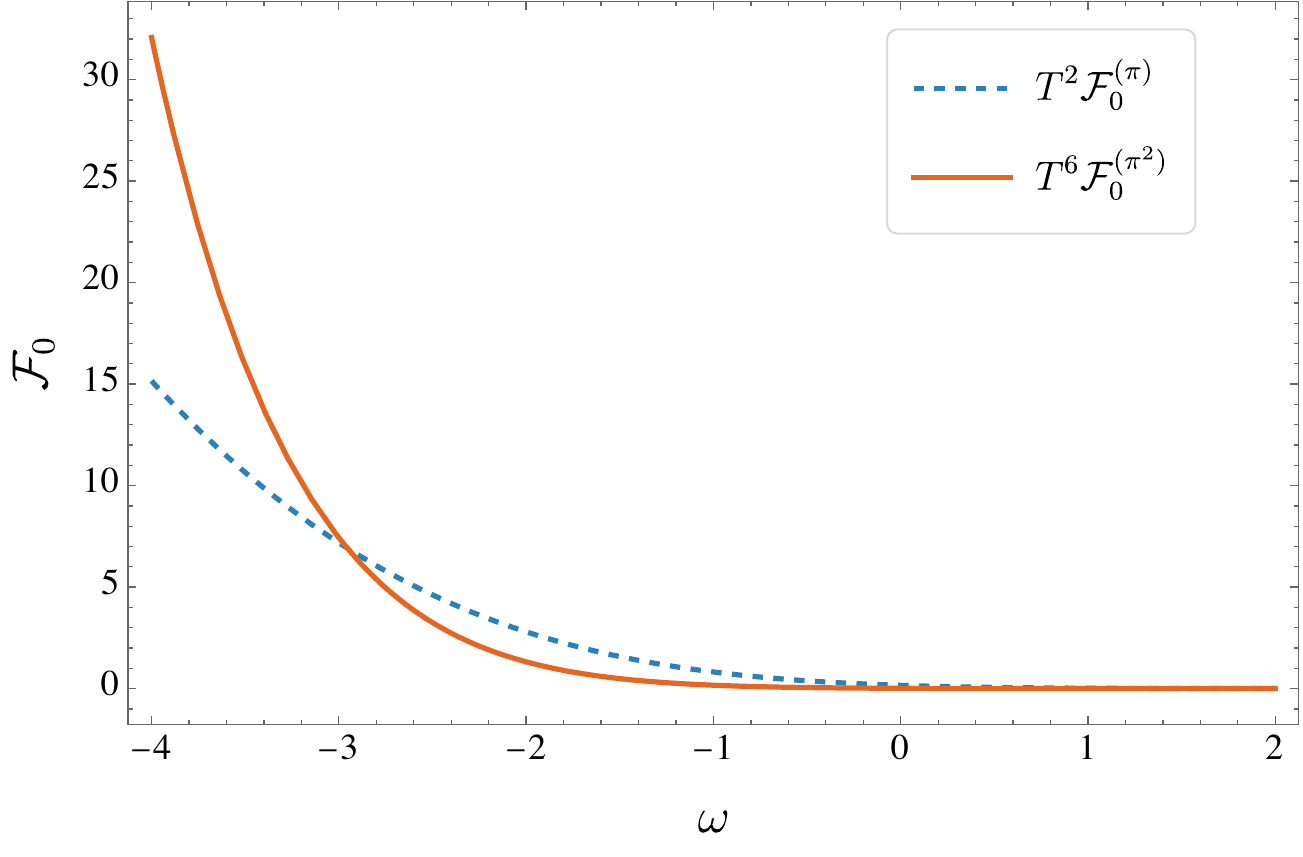}
   \caption{Response functions in the Minkowski vacuum for a Gaussian switching, shown as functions of $\omega=\Omega T$. The solid curve correspondes to the dimensionless quantity $T^2\mathcal{F}^{(\pi)}_{0}$, and the dashed curve to the numerical result for $T^6\mathcal{F}^{(\pi^2)}_{0}$, which is also dimensionless. Example in $3 + 1$ dimensions.}
    \label{fig:example_3_1_Mink}
\end{figure}

Once more, let us analyze an example using the Gaussian switching of Eq.~\eqref{eq:gaussian_switching}, where the response function $\mathcal{F}^{(\pi^2)}_{0}$ will be compared with $\mathcal{F}^{(\pi)}_{0}$. In $3 + 1$ dimensions, Eq.~\eqref{eq:response_derivative_coupling_Mink} yields
\begin{align}
    \mathcal{F}^{(\pi)}_{0} & = \frac{1}{4\pi^2}\left[
    \ii\pi\,\Phi^{(3)}(0)
    + \operatorname{PV}\int_{-\infty}^{\infty}\dd u \,\,
    \frac{\Phi^{(3)}(u)}{u}
    \right] \nonumber \\ & = -\frac{\Omega^3}{4\pi}
    \int_{-\infty}^{\infty}\dd s \,\chi(s)^2
    -\frac{3\Omega}{4\pi}
    \int_{-\infty}^{\infty}\dd s \,\chi'(s)^2
    \nonumber\\
    &\quad
    +\frac{1}{4\pi^2}\int_0^\infty \dd u \,\,
    \frac{\Phi^{(3)}(u)-\Phi^{(3)}(-u)}{u},
\end{align}
which is, up to a multiplicative factor of $1/3$, the same response function of the quadratic momentum coupling in $1 + 1$ dimensions, Eq.~\eqref{eq:pi2_mink_response_1_plus_1}. Thus, in the particular case of a Gaussian switching, we have (with $\omega = \Omega T$),
\begin{equation}
    \mathcal{F}^{(\pi)}_{0} = \frac{1}{T^2} \frac{2e^{-\frac{\omega^2}{2}}\left(2+\omega^2\right)-\sqrt{2\pi}\,\omega\left(3+\omega^2\right)\operatorname{erfc}\!\left(\frac{\omega}{\sqrt{2}}\right)}{8\pi}.
\end{equation}
The response function of Eq.~\eqref{eq:response_derivative_67} can be evaluated numerically. In Fig.~\ref{fig:example_3_1_Mink}, we display the results as a function of $\omega = \Omega T$. Qualitatively, the behavior of the response functions is similar to the $1 + 1$ case: for positive energy gaps, the momentum coupling model is more prone to excitation, whereas the magnitude of $\mathcal{F}^{(\pi^2)}_{0}$ is larger than $\mathcal{F}^{(\pi)}_{0}$ for large negative values of the energy gap (in this particular case, $\omega \lesssim - 3$).

\section{Field prepared in the firewall state - finite response}
\label{sec:firewall}

In this section, we review the main results presented in Ref.~\cite{Louko2014}. In~\cite{Louko2014}  two interaction models are considered in $1+1$ Minkowski spacetime: the standard UDW model, where the detector linearly couples to the field amplitude $\hat{\phi}$, and the derivative-coupling model, where the detector couples to its canonical momentum $\hat{\pi} = \partial_{\tau}\hat{\phi}$. In both situations, the response of the detector when crossing the Rindler firewall is shown to be finite. Let us first revisit the derivative-coupling calculation---performed in \cite{Louko2014} using integration-by-parts techniques---but using the distributional methods that will be employed throughout the remainder of this paper. This is essential to compare with the non-linear coupling case, where an integration-by-parts treatment is no longer available in a straightforward manner, and a distributional analysis is necessary.
\begin{figure}[t]
\centering
\includegraphics[width=0.8\linewidth]{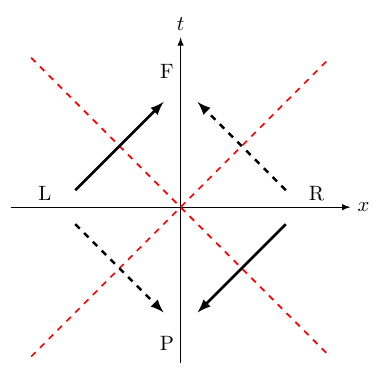}
\caption{Propagation of the decoupled sectors of a massless scalar field. The right-moving sector propagates from L into F and from R into P, while the left-moving sector propagates from R into F and from L into P}
\label{fig:firewall_extension}
\end{figure}

\subsection{Construction of the firewall state}

Let us start by recalling how the firewall state, $\hat{\rho}_{\tc{fw}}$, is defined. The goal is to construct a state that is locally indistinguishable from the Minkowski vacuum, but whose correlations between the right (R) and left (L) wedges of the Rindler spacetime have been severed. By restricting the Minkowski vacuum to the right wedge, one obtains a thermal state $\hat{\rho}_{\tc{r}}$ with respect to the Killing vector $\xi  = x \partial_{t} + t \partial_{x}$ \cite{Unruh1976, matsasUnruh}. Likewise, the restriction of the vacuum to the left wedge yields a thermal state $\hat{\rho}_{\tc{l}}$, at the same temperature as $\hat{\rho}_{\tc{r}}$, with respect to the Killing vector $-\xi$. Now, on the set $L \cup R$, we define the mixed state
\begin{equation}
    \hat{\rho}_{\tc{fw}} \equiv \hat{\rho}_{\tc{l}} \otimes \hat{\rho}_{\tc{r}}.
    \label{eq:firewallstatedef}
\end{equation}
Let us see how this definition translates into the characterization of the Wightman function
\begin{equation}
    W_{\tc{fw}}(\mf x, \mf x') = \text{Tr}\left[\hat{\phi}(\mf x)\hat{\phi}(\mf x')\hat{\rho}_{\tc{fw}}\right],
\label{eq:wightman_firewall}
\end{equation}
It follows directly from the definition that, whenever the spacetime points $\mf x$ and $\mf x'$ are in the same wedge, the firewall Wightman function $W_{\tc{fw}}$ is indistinguishable from the vacuum Wightman,
\begin{align}
     W_{\tc{fw}}(\mf x, \mf x') = \, &
    W_{0}(\mf x, \mf x'),
    \nonumber \\ 
    & \text{for } (\mf x,\mf x') \in L\times L \text{ or } R\times R.
    \label{eq:Wfwsame}
\end{align}
If \(\mf x \in R\) and \(\mf x' \in L\), the field operators are localized in different wedges, and therefore can be written as
\begin{equation}
    \hat{\phi}(\mf x) = \hat{\phi}_{\tc{r}}(\mf x)\otimes \openone_{\tc{l}},
    \qquad
    \hat{\phi}(\mf x') = \openone_{\tc{r}}\otimes \hat{\phi}_{\tc{l}}(\mf x').
\end{equation}
Then, the Wightman function becomes
\begin{align}
    W_{\tc{fw}}(\mf x, \mf x')
    &= \Tr\!\left[
    (\hat{\phi}_{\tc{r}}(\mf x)\otimes \openone_{\tc{l}})
(\openone_{\tc{r}}\otimes \hat{\phi}_{\tc{l}}(\mf x'))
(\hat{\rho}_{\tc{r}}\otimes \hat{\rho}_{\tc{l}})
    \right] \nonumber \\
    &= \Tr_{\tc{r}}\!\left[\hat{\phi}_{\tc{r}}(\mf x)\hat{\rho}_{\tc{r}}\right]
\Tr_{\tc{l}}\!\left[\hat{\phi}_{\tc{l}}(\mf x')\hat{\rho}_{\tc{l}}\right].
\end{align}
Now, since $\hat{\rho}_{\tc{r}}$ and $\hat{\rho}_{\tc{l}}$ are thermal states, their one-point functions vanish,
\begin{equation}
    \Tr_{\tc{r}}\!\left[\hat{\phi}_{\tc{r}}(\mf x)\hat{\rho}_{\tc{r}}\right] = 0,
    \qquad
    \Tr_{\tc{l}}\!\left[\hat{\phi}_{\tc{l}}(\mf x')\hat{\rho}_{\tc{l}}\right] = 0.
\end{equation}
Hence,
\begin{equation}
    W_{\tc{fw}}(\mf x, \mf x') = 0,
    \qquad
    \mf x \in R,\ \mf x' \in L.
    \label{eq:severance1}
\end{equation}
By the same reasoning,
\begin{equation}
    W_{\tc{fw}}(\mf x, \mf x') = 0,
    \qquad
    \mf x \in L,\ \mf x' \in R.
    \label{eq:severance2}
\end{equation}
So the definition of the firewall state as in Eq.~\eqref{eq:firewallstatedef} indeed accomplishes the goal: there are no correlations between the left and right wedges, and locally, the correlations are the same as in the vacuum.

Now, we would like to extend the firewall to the other wedges. As argued in Ref.~\cite{Louko2014}, the factorization of the massless 1+1-dimensional field into independent left and right moving sectors allows one to extend $\hat{\rho}_{\tc{fw}}$ from $L\cup R$ to the whole spacetime. Imposing the additional requirement that no distributional contribution be supported on the Rindler horizon then selects a unique extension. To construct such an extension, first observe that the massless vacuum Wightman, Eq.~\eqref{eq:Wightman_IR}, can be rewritten in terms of the light-cone coordinates $U = t - x$ and $V = t + x$ as
\begin{equation}
   W_{0}(\mf x, \mf x') = -\frac{1}{4\pi} \log[\Lambda(\epsilon + \ii \Delta U)] - \frac{1}{4\pi} \log[\Lambda(\epsilon + \ii \Delta V)],
\end{equation}
with $\Delta U = U - U'$ and $\Delta V = V - V'$, up to a real-valued additive constant, which we may absorb in $\Lambda$. Moreover, the logarithms are considered in their principal branches \footnote{This will be the case in all the formulas that involve the complex logarithm in this paper.}. This shows that the right-moving ($U$) and left-moving ($V$) sectors decouple. Consequently, the left-moving sector propagates into the future wedge F only from R, and into the past wedge P only from L, while the right-moving sector propagates into F only from L, and into P only from R (see Fig.~\ref{fig:firewall_extension}). This determines a unique extension of the state to $\text{F} \cup \text{P} \cup \text{R} \cup \text{L}$. To extend the corresponding Wightman function to the whole spacetime, one must specify its behavior on the Rindler horizon.  As in Ref.~\cite{Louko2014}, we choose the minimal extension, namely, the one that introduces no distributional contribution supported on the horizon. This prescription yields a unique continuation, which we denote simply by $W_{\tc{fw}}(\mf x, \mf x')$. Evidently, the properties specified in Eq.~\eqref{eq:Wfwsame}, \eqref{eq:severance1}, and \eqref{eq:severance2} still hold. 

For the quadratic-operator analysis below, we take the extended firewall state to be Gaussian with vanishing one-point function and two-point function $W_{\tc{fw}}$. Its higher $n$-point functions are therefore fixed by Wick factorization. 

As for the wedges split by the Rindler horizon, we have
\begin{align}
    W_{\tc{fw}}(\mf x, \mf x')
    &= -\frac{1}{4\pi}\log\!\bigl[\Lambda\bigl(\epsilon + \ii
    \Delta U\bigr)\bigr],
    \nonumber\\
    &\hphantom{={}}
    \text{for } (\mf x,\mf x') \in \text{P}\times \text{R}
    \text{ or } \text{L}\times \text{F} .
    \label{eq:Wightman_firewall_rel1}\\
    W_{\tc{fw}}(\mf x, \mf x')
    &= -\frac{1}{4\pi}\log\!\bigl[\Lambda\bigl(\epsilon + \ii\Delta V\bigr)\bigr],
    \nonumber\\
    &\hphantom{={}}
    \text{for } (\mf x,\mf x') \in \text{P}\times \text{L}
    \text{ or } \text{R}\times \text{F} .
    \label{eq:Wightman_firewall_rel2}
\end{align}

\begin{figure}[!h]
    \centering
    \includegraphics[width=0.8\linewidth]{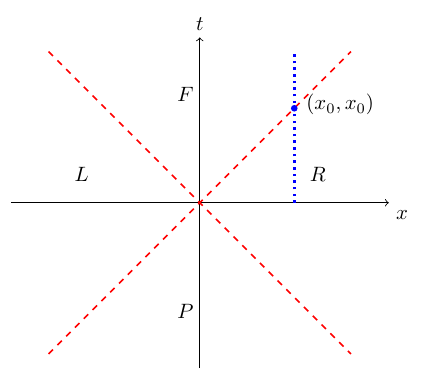}
    \caption{Schematic representation of an inertial pointlike detector (dotted line) crossing the Rindler firewall (dashed line) from the right (R) and future (F) wedges.}
    \label{fig:wedges}
\end{figure}

\subsection{Firewall response in the derivative coupling model}

As before, we assume that the detector starts in the ground state, so that the initial state of the full system is
\begin{equation}
    \hat{\rho}_{0} = |g \rangle \langle g| \otimes \hat{\rho}_{\tc{fw}}.
\end{equation}
We work in $1 + 1$ Minkowski spacetime with inertial coordinates $\mf x = (t, x)$, assuming a pointlike detector with spacetime smearing $\Lambda(t, x)$ given by Eq.~\eqref{eq:pointlike_smearing}. Here, $(t, x)= (x_{0}, x_{0})$ are the coordinates of the spacetime point where the detector crosses the Rindler horizon. We shall assume $\chi(t)$ to be compactly supported on the interval $[0, T]$, with $T > x_{0}$ (see Fig.~\ref{fig:wedges}). The switching function also has to be smooth enough, so we can prevent any divergences arising from switching artifacts.

The detector couples to the momentum of the field according to the  Hamiltonian density \eqref{eq:derivativeHamo}. For a pointlike detector, from Eq.~\eqref{eq:response_derivative_coupling_Mink} the detector's response function reads
\begin{equation}
\mathcal{F}^{(\pi)}_{\tc{fw}} = \int_{-\infty}^{\infty} \dd t \, \int_{-\infty}^{\infty} \dd t \,' \chi(t)\chi(t')e^{-\ii \Omega (t - t')}\partial_{t}\partial_{t'}W_{\tc{fw}}(t, t'),
\end{equation}
where $W_{\tc{fw}}(t, t')$ is the two-point function pulled back to the detector's trajectory and the firewall Wightman is given by \eqref{eq:Wightman_firewall_rel1}, \eqref{eq:Wightman_firewall_rel2}.  Let us now show how $\partial_{t}\partial_{t'}W_{\tc{fw}}(t, t')$ can be explicitly written in terms of well-defined distributions.

First, we need to write a full expression for $W_{\tc{fw}}(\mf x, \mf x')$ (restricted to the R-F crossing). To that end, let us define the indicator functions of the wedges R and F. Using standard inertial coordinates $\mf x = (t, x)$, we define the following
\begin{equation}
    \mathds{1}_{\tc{r}}(t, x) = \Theta(x)\Theta(x^2 - t^2),
\end{equation}
\begin{equation}
    \mathds{1}_{\tc{f}}(t, x) =  \Theta(t)\Theta(t^2 - x^2).
\end{equation}
We also define the following auxiliary functions:
\begin{equation}
    f_{0}(\mf x, \mf x')  = \mathds{1}_{\tc{r}}(t, x)\mathds{1}_{\tc{r}}(t', x') + \mathds{1}_{\tc{f}}(t, x)\mathds{1}_{\tc{f}}(t', x'),
\end{equation}
\begin{equation}
    f_{\tc{rf}}(\mf x, \mf x')  = \mathds{1}_{\tc{r}}(t, x)\mathds{1}_{\tc{f}}(t', x'), 
\end{equation}
\begin{equation}
    f_{\tc{fr}}(\mf x, \mf x')  = \mathds{1}_{\tc{f}}(t, x)\mathds{1}_{\tc{r}}(t', x'). 
\end{equation}
It will be helpful for later sections to introduce two auxiliary distributions $W_{\tc{rf}}$ and $W_{\tc{fr}}$, defined on spacetime pairs, which coincide with the firewall Wightman function whenever $(\mf x,\mf x')\in R\times F$ and $F\times R$, respectively. Then we can write 
\begin{align}
    W_{\tc{fw}}(\mf x, \mf x')  
    & = f_{0}(\mf x, \mf x')W_{0}(\mf x, \mf x') \nonumber \\ & + f_{\tc{rf}}(\mf x, \mf x')W_{\tc{rf}}(\mf x, \mf x') \nonumber \\ &  +  f_{\tc{fr}}(\mf x, \mf x')W_{\tc{fr}}(\mf x, \mf x').
    \label{eq:full_WFW}
\end{align}
Next, because $f_{0}(\mf x, \mf x') = f_{0}(\mf x', \mf x)$ and $f_{\tc{rf}}(\mf x', \mf x) = f_{\tc{fr}}(\mf x, \mf x')$, we have
\begin{align}
      W_{\tc{fw}}(\mf x', \mf x)  &= f_{0}(\mf x', \mf x)W_{0}(\mf x', \mf x) + f_{\tc{rf}}(\mf x', \mf x)W_{\tc{rf}}(\mf x', \mf x) \nonumber \\ & +  f_{\tc{fr}}(\mf x', \mf x)W_{\tc{fr}}(\mf x', \mf x) \nonumber \\ 
      & = f_{0}(\mf x, \mf x')W_{0}(\mf x', \mf x) + f_{\tc{fr}}(\mf x, \mf x')W_{\tc{rf}}(\mf x', \mf x) \nonumber \\ & +  f_{\tc{rf}}(\mf x, \mf x')W_{\tc{fr}}(\mf x', \mf x).
      \label{eq:Wfw_xprimex}
\end{align}
Since $W_{\tc{fw}}(\mf x, \mf x')$ is a Wightman function, the following must hold,
\begin{equation}
    W_{\tc{fw}}(\mf x', \mf x) = W_{\tc{fw}}(\mf x, \mf x')^{*}.
    \label{eq:complexsymmetry}
\end{equation}
Thus, taking the complex conjugate of Eq.~\eqref{eq:full_WFW} and using Eq.~\eqref{eq:Wfw_xprimex}, it follows that
\begin{align}
    & f_{0}(\mf x, \mf x')W_{0}(\mf x, \mf x')^{*} + f_{\tc{rf}}(\mf x, \mf x')W_{\tc{rf}}(\mf x, \mf x')^{*} \nonumber \\ & +  f_{\tc{fr}}(\mf x, \mf x')W_{\tc{fr}}(\mf x, \mf x')^{*} \nonumber \\ 
    & = f_{0}(\mf x, \mf x')W_{0}(\mf x', \mf x) + f_{\tc{fr}}(\mf x, \mf x')W_{\tc{rf}}(\mf x', \mf x) \nonumber \\ & +  f_{\tc{rf}}(\mf x, \mf x')W_{\tc{fr}}(\mf x', \mf x).
    \label{eq:anotherstepp}
\end{align}
By Eqs.~\eqref{eq:Wightman_IR} and \eqref{eq:Wightman_firewall_rel2}, it is clear that $W_{0}(\mf x', \mf x) = W_{0}(\mf x, \mf x')^{*}$ and $W_{\tc{rf}}(\mf x', \mf x) = W_{\tc{rf}}(\mf x, \mf x')^{*}$. Thus, Eq.~\eqref{eq:anotherstepp} simplifies to
\begin{align}
    & W_{\tc{rf}}(\mf x, \mf x')^{*}(f_{\tc{rf}}(\mf x, \mf x') - f_{\tc{fr}}(\mf x, \mf x'))  = \nonumber \\ & f_{\tc{rf}}(\mf x, \mf x')W_{\tc{fr}}(\mf x', \mf x) - f_{\tc{fr}}(\mf x, \mf x')W_{\tc{fr}}(\mf x, \mf x')^{*}
\end{align}
If $(\mf x, \mf x') \in (\text{R}, \text{F})$, we obtain
\begin{equation}
    W_{\tc{rf}}(\mf x, \mf x')^{*} = W_{\tc{fr}}(\mf x', \mf x),
    \label{eq:relation1}
\end{equation}
whereas for $(\mf x, \mf x') \in (\text{F}, \text{R})$,
\begin{equation}
    W_{\tc{rf}}(\mf x, \mf x')^{*} = W_{\tc{fr}}(\mf x, \mf x')^{*}
    \label{eq:relation2}
\end{equation}
Given Eqs.~\eqref{eq:relation1} and \eqref{eq:relation2}, we conclude that
\begin{equation}
    W_{\tc{rf}}(\mf x, \mf x') = W_{\tc{fr}}(\mf x, \mf x').
\end{equation}
Let us re-parameterize the trajectory to simplify the upcoming calculations. Let $X = x - x_{0}$ and $\tau = t - x_{0}$. In the coordinates $(\tau, X)$, the detector follows the trajectory $X(\tau) = 0$, and, since $x_0>0$, the indicator functions of the wedges R and F pulled back to the detector's trajectory become
\begin{equation}
    \mathds{1}_{\tc{r}}(\tau) = \Theta(-\tau)\Theta(\tau + 2 x_{0}), 
\end{equation}
and
\begin{equation}
    \mathds{1}_{\tc{f}}(\tau)=\Theta(\tau).
\end{equation}
Then, we can write
\begin{align} \label{eq:full_WFW_pullback}
     W_{\tc{fw}}(\tau,  \tau')
     & = -\frac{1}{4\pi}\log[-\Lambda^2(\Delta \tau - \ii \epsilon)^2]\; f_{0}(\tau, \tau')
     \\* &  - \frac{1}{4\pi}\log[\Lambda(\epsilon + \ii \Delta \tau)]\;\bigl(f_{\tc{rf}}(\tau, \tau') + f_{\tc{fr}}(\tau, \tau')\bigr).\nonumber
\end{align}
where, on the detector's worldline,
\begin{align}
    f_{0}(\tau,\tau') &= \Theta(-\tau)\Theta(-\tau')\Theta(\tau + 2 x_{0})\Theta(\tau' + 2 x_{0}) \label{eq:f_0} \nonumber \\ & +\Theta(\tau)\Theta(\tau'), \\
    f_{\tc{rf}}(\tau,\tau') &= \Theta(-\tau)\Theta(\tau + 2 x_{0})\Theta(\tau'), \\
    f_{\tc{fr}}(\tau,\tau') &= \Theta(\tau)\Theta(-\tau')\Theta(\tau' + 2x_{0}).
    \label{eq:f_auxiliary_tau_tau'}
\end{align}
Thus, 
\begin{align}
\partial_{\tau}\partial_{\tau'}W_{\tc{fw}}(\tau,\tau')
= -\frac{1}{4\pi}\Bigl[
&\;\partial_{\tau}\partial_{\tau'}\!\bigl(f_{0}(\tau,\tau')\,L_{0}(\Delta\tau)\bigr)
\nonumber \\ & +\partial_{\tau}\partial_{\tau'}\!\bigl(f_{\tc{rf}}(\tau,\tau')\,L_{1}(\Delta\tau)\bigr)
+ \nonumber \\ & \partial_{\tau}\partial_{\tau'}\!\bigl(f_{\tc{fr}}(\tau,\tau')\,L_{1}(\Delta\tau)\bigr)
\Bigr],
\label{eq:pulledbackWFW}
\end{align}
where
\begin{equation}
    L_{0}(\Delta\tau)=\log\!\bigl[-\Lambda^{2}(\Delta\tau-\ii\epsilon)^{2}\bigr],
    \label{eq:defL0}
\end{equation}
and
\begin{equation}
    L_{1}(\Delta\tau)=\log\!\bigl[\Lambda(\epsilon+\ii\Delta\tau)\bigr].
    \label{eq:defL1}
\end{equation}
By evaluating the derivatives in Eq.~\eqref{eq:pulledbackWFW}, we can split the response function into three contributions, each to be treated distributionally in the limit $\epsilon \to 0^{+}$. Concretely,
\begin{equation}
    \mathcal{F}_{\tc{fw}}^{(\pi)} = -\frac{1}{2\pi} \lim_{\epsilon \to 0^{+}}(\mathcal{A} + \mathcal{B} + \mathcal{C}),
\end{equation}
where (with $g(\tau, \tau') \equiv \chi(\tau)\chi(\tau')e^{-\ii \Omega \Delta \tau}$)
\begin{align}
    \mathcal{A} & \equiv \int_{-\infty}^{\infty}\int_{-\infty}^{\infty} \dd \tau \, \dd \tau \,' \frac{g(\tau, \tau')}{(\Delta \tau - \ii \epsilon)^2}\Biggl[f_{0}(\tau, \tau') \nonumber \\ & + \frac{f_{\tc{rf}}(\tau, \tau') + f_{\tc{fr}}(\tau, \tau')}{2} \Biggr],
    \label{eq:mathcalA_main_text}
\end{align}
\begin{align}
\mathcal{B} \equiv {} &- \int_{-\infty}^{\infty}\int_{-\infty}^{\infty}
\dd \tau \, \dd \tau \,'\,
\frac{g(\tau, \tau')}{\Delta \tau - \ii \epsilon}
\Biggl[
\partial_{\tau}f_{0}(\tau, \tau')
- \partial_{\tau'}f_{0}(\tau, \tau')
\nonumber\\
&\quad
+ \frac{
\partial_{\tau}f_{\tc{rf}}(\tau, \tau')
+ \partial_{\tau}f_{\tc{fr}}(\tau, \tau')
}{
2}
\nonumber\\
&\quad
- \frac{
\partial_{\tau'}f_{\tc{rf}}(\tau, \tau')
+ \partial_{\tau'}f_{\tc{fr}}(\tau, \tau')
}{2}
\Biggr],
\label{eq:term_mathcal_B_main_text}
\end{align}
and
\begin{align}
    \mathcal{C} & \equiv \frac{1}{2}\int_{-\infty}^{\infty}\int_{-\infty}^{\infty} \dd \tau \, \dd \tau \,' g(\tau, \tau')\Biggl[ L_{0}(\Delta \tau) \partial_{\tau}\partial_{\tau'}f_{0}(\tau, \tau') \nonumber \\ & +  L_{1}(\Delta \tau) \partial_{\tau}\partial_{\tau'}f_{\tc{rf}}(\tau, \tau') +  L_{1}(\Delta \tau) \partial_{\tau}\partial_{\tau'}f_{\tc{fr}}(\tau, \tau')\Biggr].
    \label{eq:term_mathcal_C_main_text}
\end{align}
The evaluation of each of those terms, and consequently of the response function, is presented in Appendix \ref{sec:appendix_distributions_firewall}. There, we develop extensions of the identity \eqref{eq:generalizedSP_main_text} to the half-line. It is interesting to note that each of the terms $\mathcal{A}$, $\mathcal{B}$, and $\mathcal{C}$ contains persistent divergences in the limit $\epsilon \to 0^{+}$ arising from the singularities at the origin when evaluating integrals in the half-line. Nonetheless, those divergences cancel out perfectly, yielding a finite response function $\mathcal{F}^{(\pi)}_{\tc{fw}}$. The result, which was previously obtained by~\cite{Louko2014} via integration by parts, can be written in terms of the vacuum response function, $\mathcal{F}^{(\pi)}_{0}$. Using exactly the same procedure as in Sec.~\ref{sec:Mink1_1}, together with identity \eqref{eq:generalizedSP_main_text}, we have
\begin{align}
    \mathcal{F}^{(\pi)}_{0} & = -\frac{1}{2\pi} \lim_{\epsilon \to 0^{+}} \int_{-\infty}^{\infty}\dd u \, \frac{\Phi(u)}{(u - \ii \epsilon)^2} \nonumber \\ & = -\frac{1}{2\pi}\Biggl[\ii \pi \Phi'(0) + \operatorname{PV}\int_{-\infty}^{\infty} \dd u \, \frac{\Phi'(u)}{u} \Biggr] \nonumber \\ & = -\frac{\Omega}{2}\int_{-\infty}^{\infty}\dd s\,\chi(s)^2
    - \frac{1}{2\pi}\operatorname{PV}\int_{-\infty}^{\infty}\dd u \,\,
    \frac{\Phi'(u)}{u}.
\end{align}
The result is then
\begin{align}
    \mathcal{F}^{(\pi)}_{\tc{fw}} & =  \mathcal{F}^{(\pi)}_{0} + \frac{\chi(0)^2}{2\pi}\log(|\Omega|e^{\gamma - 1}/\Lambda) \nonumber \\ & +  \frac{1}{2\pi}\int_{0}^{\infty}\dd u \, \cos(\Omega u)\Biggl[\frac{\chi(0)(\chi(0) - \chi(u) - \chi(-u))}{u}  \nonumber \\ & + \frac{1}{u^2} \int_{0}^{u}\dd s \chi(s)\chi(s - u)\Biggr],
    \label{eq:requiem_main_text}
\end{align}
where $\gamma$ is the Euler--Mascheroni constant. This expression shows that the firewall response retains the standard infrared ambiguity of the massless scalar field in $1+1$ dimensions: it depends on the infrared scale $\Lambda$, with a logarithmic divergence as $\Lambda \to 0$. This reflects the familiar IR ambiguity  of the $1+1$ scalar field, whose Wightman function is only defined after aa small mass/IR scale has been fixed~\cite{birrell_davies}.

\begin{figure}
    \centering
    \includegraphics[width=1.0\linewidth]{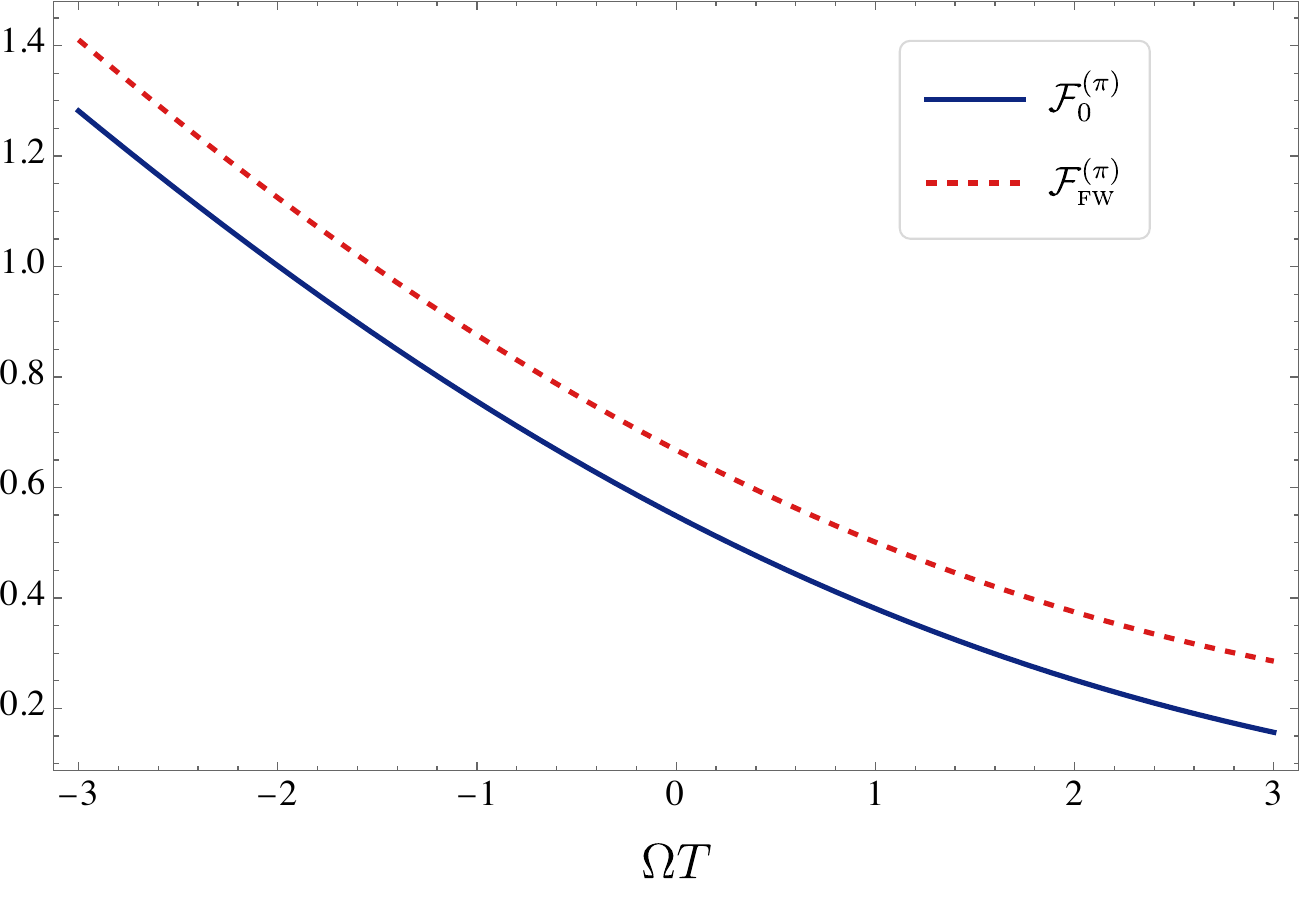}
    \caption{Comparison between the detector response when the field is prepared in the vacuum, $\mathcal{F}_{0}^{(\pi)}$, and when the field is prepared in the firewall state, $\mathcal{F}_{\tc{fw}}^{(\pi)}$, in the derivative-coupling model, as a function of the dimensionless energy gap $\Omega T$. The Rindler horizon crossing occurs at $t = x_{0} = 0.5\,T$, and the IR cutoff is fixed at $\Lambda T = 1$.}
\label{fig:comparison1}
\end{figure}

To further compare the qualitative behavior of the response functions $\mathcal{F}^{(\pi)}_{\tc{fw}}$ and $\mathcal{F}^{(\pi)}_{0}$, we choose the switching (expressed in the shifted coordinate $\tau = t - x_{0}$)
\begin{equation}
\chi(\tau)=
\begin{cases}
\sin^{2}\!\left(\dfrac{\pi(\tau+x_{0})}{T}\right), & \tau\in[-x_{0},\,T-x_{0}],\\[6pt]
0, & \text{otherwise},
\end{cases}
\end{equation}
where we recall that $t = x_{0}$ corresponds to the Rindler horizon crossing (see Fig.~\ref{fig:wedges}), with $T > x_{0}$. In Fig.~\ref{fig:comparison1}, we plot both response functions as functions of the scaled energy gap $\Omega T$, fixing $x_{0} = 0.5 T$ and $\Lambda T = 1$. We see that the firewall response has the same qualitative behavior as the vacuum response, but is systematically larger throughout the range shown. Thus, in this case, the net effect of the horizon crossing, when the field is prepared in the firewall state, is to enhance the detector response.

\begin{figure}
    \centering
    \includegraphics[width=1.0\linewidth]{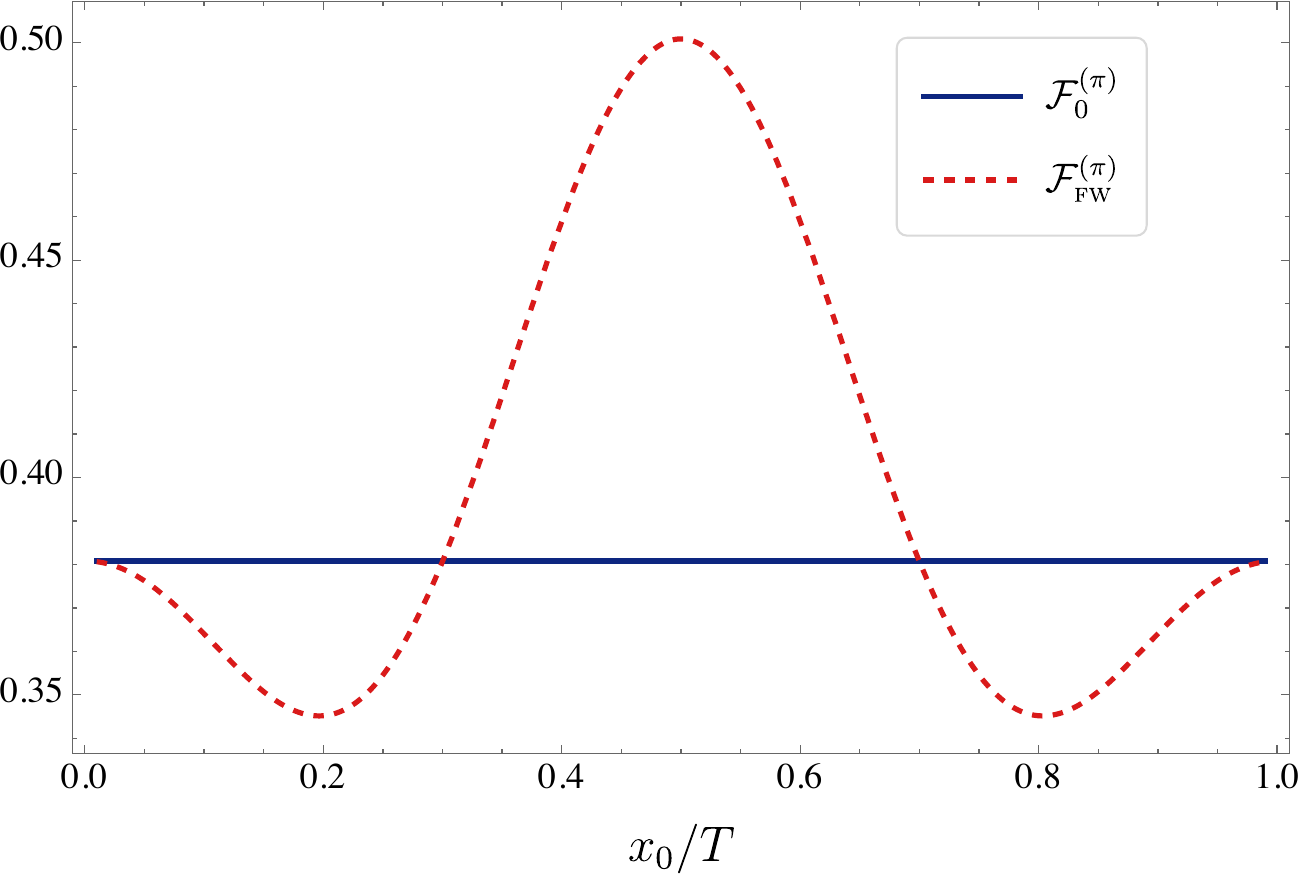}
    \caption{Comparison between the detector response when the field is prepared in the vacuum, $\mathcal{F}_{0}^{(\pi)}$, and when the field is prepared in the firewall state, $\mathcal{F}_{\tc{fw}}^{(\pi)}$, in the derivative-coupling model, as a function of the dimensionless Rindler horizon crossing coordinate time, $x_{0}/T$. The detector's gap is fixed at $\Omega T = 1$, and the IR cutoff is $\Lambda T = 1$.}
\label{fig:comparison2}
\end{figure}

This enhancement, however, generally depends on the relative timing between the switching and the horizon crossing, as illustrated in Fig.~\ref{fig:comparison2}. In that figure, we fix $\Omega T = \Lambda T = 1$ and vary the horizon-crossing time $x_{0}$. When $x_{0} = 0.5 T$, the enhancement relative to the vacuum case is maximal. On the other hand, there are two values of $x_{0}$, approximately $0.3 T$ and $0.7 T$, for which the two response functions coincide exactly. We also observe two regions near the endpoints where $\mathcal{F}^{(\pi)}_{\tc{fw}}$ is actually smaller than $\mathcal{F}^{(\pi)}_{0}$. This is remarkable: depending on switching timing, crossing the firewall can both make the detector more excited or calm it down, reducing its baseline vacuum excitation.

\begin{figure}
    \centering
    \includegraphics[width=1.0\linewidth]{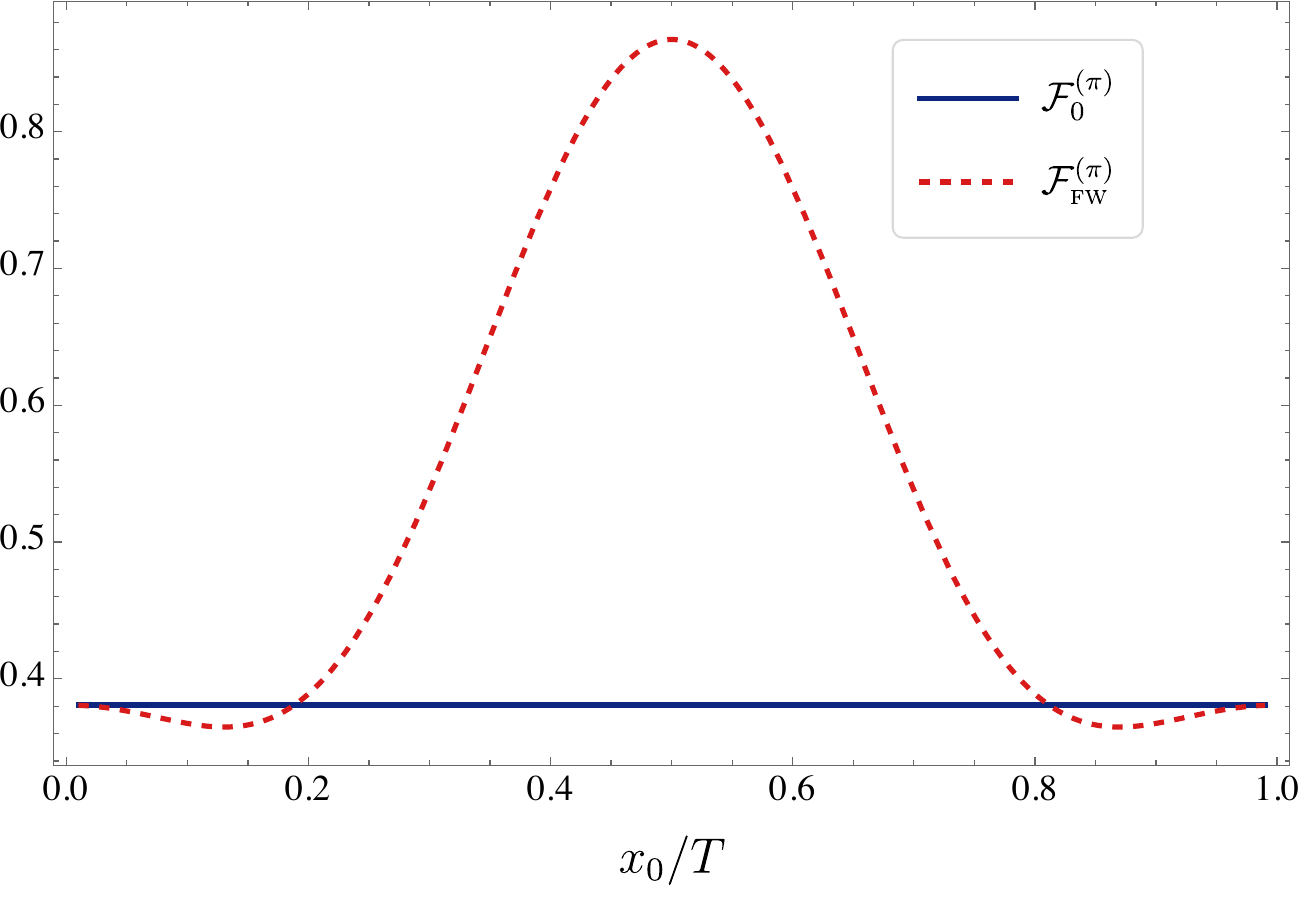}
    \caption{Comparison between the detector response when the field is prepared in the vacuum, $\mathcal{F}_{0}^{(\pi)}$, and when the field is prepared in the firewall state, $\mathcal{F}_{\tc{fw}}^{(\pi)}$, in the derivative-coupling model, as a function of the dimensionless Rindler horizon crossing coordinate time, $x_{0}/T$. The detector's gap is fixed at $\Omega T = 1$, and the IR cutoff is $\Lambda T = 0.1$.}
    \label{fig:comparison2_anotherLambda}
\end{figure}

As $x_{0} \to 0$ or $x_{0} \to T$, the two response functions coincide, as expected, since in these limits the detector couples to the field only on one side of the Rindler horizon, where the firewall state is indistinguishable from the vacuum. The same qualitative behavior is observed for a smaller value of the IR cutoff, for example $\Lambda T = 0.1$, although in that case the region where $\mathcal{F}^{(\pi)}_{\tc{fw}} < \mathcal{F}^{(\pi)}_{0}$ is less pronounced (see Fig.~\ref{fig:comparison2_anotherLambda}).

\section{The obstruction to a well-defined quadratic response}
\label{sec:andthenitdiverges}

As we reviewed in Sec.~\ref{sec:firewall}, the detector response in the firewall state is finite when the detector couples linearly either to the field operator $\hat{\phi}$ or to its conjugate momentum $\hat{\pi}$ \cite{Louko2014}. A natural follow-up question is what happens for detector models coupled to composite operators, in particular to the energy density of the field. Since the firewall state is not Hadamard \cite{fewsterNecessityHadamard, Louko2014, EduJormaPRLFire}, additional singular behavior may arise in this case. A direct evaluation of the corresponding response function would, however, be technically cumbersome. Motivated by our previous analysis of pointlike detectors in Sec.~\ref{sec:Minkowski_vacuum}, where quadratic momentum coupling provides a useful proxy for energy-density coupling in the Minkowski vacuum, we focus on this simpler model. Nonetheless, we do not assume that this vacuum correspondence persists in the firewall state.

Before proceeding, it is important to clarify the role of normal ordering in the analysis below. The physical interaction model considered in this work remains the one introduced in Sec.~\ref{sec:Minkowski_vacuum}, in which $\hat{\pi}^{2}$ is normal ordered with respect to the Minkowski vacuum, as in Eq.~\eqref{eq:normal_ordering_standard}. This vacuum subtraction is part of the definition of the interaction; it is not assumed to render composite-observable correlations well defined in the non-Hadamard firewall state. In particular, vacuum normal ordering does not by itself guarantee that the complete quadratic-observable kernel
\begin{equation}
\operatorname{Tr}\!\left[
\mathopen{:}\hat{\pi}^{2}(\mf x)\mathclose{:}
\mathopen{:}\hat{\pi}^{2}(\mf x')\mathclose{:}
\hat{\rho}_{\tc{fw}}
\right]
\end{equation}
admits a canonical, regulator-independent definition. This is a potential obstruction to defining the detector response in the sharp-firewall model, but it does not by itself prove that no common extension or regularization of the complete kernel exists.
Nonetheless, since our goal is to isolate the features of the sharp-firewall model that could prevent the construction of a well-defined detector response, we instead use the notation of a firewall-relative subtraction (that we will refer to as firewall-state normal ordering) described in Appendix \ref{sec:normal_ordering_general}  as a formal algebraic device. If used to define the interaction Hamiltonian, changing the normal-ordering reference would in general produce a physically inequivalent detector model, since the two prescriptions differ by a reference-state-dependent $c$-number contribution.

With this understanding, the formal firewall-relative subtraction provides a conditional representation of the connected fluctuations of the operator appearing in the physical interaction. Suppose that all one- and two-point quantities constructed from the quadratic composite operator $\hat{\pi}^{2}$ in the expression below are assigned a meaning by a single common prescription, with the corresponding $c$-number subtractions treated consistently. We do not assume that this prescription is unique or that it admits a regulator-independent sharp-firewall limit. With this stipulation, one may form
\begin{equation}
\operatorname{Tr}\!\left[
\mathopen{:}\hat{\pi}^{2}(\mf x)\mathclose{:}\,
\mathopen{:}\hat{\pi}^{2}(\mf x')\mathclose{:}
\hat{\rho}_{\tc{fw}}
\right]
-
\operatorname{Tr}\!\left[
\mathopen{:}\hat{\pi}^{2}(\mf x)\mathclose{:}
\hat{\rho}_{\tc{fw}}
\right]
\operatorname{Tr}\!\left[
\mathopen{:}\hat{\pi}^{2}(\mf x')\mathclose{:}
\hat{\rho}_{\tc{fw}}
\right].
\label{eq:connected_quadratic_fw}
\end{equation}
Under this assumption, the expression above is the two-point correlation with the product of the corresponding one-point functions subtracted, which we refer to as the connected combination for the Minkowski-normal-ordered observable in the firewall state.

A connected correlation function is algebraically unchanged when the observable is shifted by a $c$-number multiple of the identity, which may depend on the spacetime point. Therefore, whenever the same prescription is used consistently for all terms, changing the normal-ordering reference amounts algebraically to the shift
\begin{equation}
\mathopen{:}\hat{\pi}^{2}(\mf x)\mathclose{:}_{\tc{fw}}
=
\mathopen{:}\hat{\pi}^{2}(\mf x)\mathclose{:}
-
\operatorname{Tr}\!\left[
\mathopen{:}\hat{\pi}^{2}(\mf x)\mathclose{:}
\hat{\rho}_{\tc{fw}}
\right]\openone .
\label{eq:normal_ordering_connected_relation}
\end{equation}
Consequently, the expression in Eq.~\eqref{eq:connected_quadratic_fw} can equivalently be represented as
\begin{align}
&\operatorname{Tr}\!\left[
\mathopen{:}\hat{\pi}^{2}(\mf x)\mathclose{:}_{\tc{fw}}
\mathopen{:}\hat{\pi}^{2}(\mf x')\mathclose{:}_{\tc{fw}}
\hat{\rho}_{\tc{fw}}
\right].
\label{eq:connected_quadratic_fw_normal_order}
\end{align}

Thus, although adopting firewall-state normal ordering in the interaction Hamiltonian would define a different physical model, we do not make such a replacement. We use the state-relative expression only as an algebraically convenient representation of the connected correlation associated with the original Minkowski-normal-ordered observable.

Using Gaussianity with zero one-point function, Wick factorization formally gives (see Appendix \ref{sec:normal_ordering_general}):
\begin{equation}
\operatorname{Tr}\!\left[
\mathopen{:}\hat{\pi}^{2}(\mf x)\mathclose{:}_{\tc{fw}}
\mathopen{:}\hat{\pi}^{2}(\mf x')\mathclose{:}_{\tc{fw}}
\hat{\rho}_{\tc{fw}}
\right]
=
2\bigl(
\partial_{\tau}\partial_{\tau'}W_{\tc{fw}}(\mf x,\mf x')
\bigr)^2.
\label{eq:connected_fw_wightman}
\end{equation}
Because the firewall state is non-Hadamard, this formal Wick factorization does not imply that the product on the right-hand side defines a bona fide distribution. Determining the distributional status of this expression is precisely the issue addressed below.

The quantity studied in the remainder of this section is therefore not the response associated with a second detector Hamiltonian: it is the formal connected-correlation response function associated with the original Minkowski-normal-ordered interaction, by which we mean the result of inserting the connected combination in Eq.~\eqref{eq:connected_quadratic_fw} into the usual detector-response integral. Whenever all the relevant terms are defined by the same prescription, the complete Minkowski-normal-ordered two-point kernel is the sum of the connected correlation and the corresponding product of one-point functions. We do not establish here that the complete kernel lacks a canonical, regulator-independent extension, or that the corresponding full detector response is undefined, since a common extension or regularization could in principle produce cancellations between these terms. Our more limited result is that the formal sharp-firewall candidate for the connected correlation contains products of distributions that are not canonically defined in standard distribution theory.

Retaining the notation $\mathcal{F}^{(\pi^2)}_{\tc{fw}}$ for the contribution obtained from this formal connected response function, and considering the same pointlike detector as in the setup of Fig.~\ref{fig:wedges}, we can formally write

\begin{equation}
    \mathcal{F}^{(\pi^2)}_{\tc{fw}}
    = 2 \int \dd \tau \, \dd \tau \,'\,
    \chi(\tau)\chi(\tau') e^{-\ii \Omega \Delta \tau}(\partial_{\tau}\partial_{\tau'}W_{\tc{fw}}(\tau, \tau'))^2,
    \label{eq:response_fw_pi2}
\end{equation}
where, as before, $\tau = t - x_{0}$.
 Before tackling any heavy calculations, let us understand the structure of this response function. To this end, we first rewrite Eq.~\eqref{eq:pulledbackWFW} as
\begin{equation}
    \partial_{\tau}\partial_{\tau'}W_{\tc{fw}}(\tau,\tau')
    = \frac{A(\tau,\tau')}{(\Delta \tau-\ii\epsilon)^2}
    + \frac{B(\tau,\tau')}{\Delta \tau-\ii\epsilon}
    + C(\tau,\tau'),
\end{equation}
where
\begin{equation}
    A(\tau, \tau')
    = -\frac{1}{4\pi}
    \bigl(
    2f_0(\tau,\tau')
    + f_{\tc{rf}}(\tau,\tau')
    + f_{\tc{fr}}(\tau,\tau')
    \bigr),
\end{equation}
\begin{align}
    B(\tau, \tau')
    = -\frac{1}{4\pi}\bigl(&
    2\partial_{\tau'}f_0(\tau,\tau')
    - 2\partial_{\tau}f_0(\tau,\tau')
    \nonumber \\ & + \partial_{\tau'}f_{\tc{rf}}(\tau,\tau')
    - \partial_{\tau}f_{\tc{rf}}(\tau,\tau')
    \nonumber\\
    &
    + \partial_{\tau'}f_{\tc{fr}}(\tau,\tau')
    - \partial_{\tau}f_{\tc{fr}}(\tau,\tau')
    \bigr),
\end{align}
and
\begin{align}
    C(\tau, \tau')
    & = -\frac{1}{4\pi}\bigl(
    L_0(\Delta \tau)\,
    \partial_{\tau}\partial_{\tau'}f_0(\tau,\tau')
    \nonumber\\
    &
    + L_1(\Delta \tau)
    \bigl(
    \partial_{\tau}\partial_{\tau'}f_{\tc{rf}}(\tau,\tau')
    + \partial_{\tau}\partial_{\tau'}f_{\tc{fr}}(\tau,\tau')
    \bigr)
    \bigr).
\end{align}
Because the switching function $\chi(\tau)$ is compactly supported in the interval $[-x_{0},\, T-x_{0}]$, with $0<x_{0}<T$, the factors $\Theta(\tau+2x_{0})$ and $\Theta(\tau'+2x_{0})$ appearing in the definitions of $f_{0}$, $f_{\mathrm{rf}}$, and $f_{\mathrm{fr}}$, Eqs.~\eqref{eq:f_0}--\eqref{eq:f_auxiliary_tau_tau'}, are identically equal to unity on the support of the integrals. Likewise, derivatives of these Heaviside functions produce Dirac deltas supported at $\tau=-2x_{0}$ or $\tau'=-2x_{0}$, which lie outside the support of the switching function and therefore give no contribution. Consequently, within the response function the auxiliary functions may be replaced by (see Appendix \ref{sec:appendix_distributions_firewall} for the explicit calculations)
\begin{equation}
    A(\tau, \tau')
   = -\frac{1}{4\pi}
    \Bigl(
    1 + \Theta(\tau)\Theta(\tau')
    + \Theta(-\tau)\Theta(-\tau')
    \Bigr),
\end{equation}
\begin{equation}
    B(\tau, \tau')
    = -\frac{1}{4\pi}
    \bigl(
    \delta(\tau')\operatorname{sgn}(\tau)
    - \delta(\tau)\operatorname{sgn}(\tau')
    \bigr),
\end{equation}
 and
\begin{equation}
    C(\tau, \tau')
    = -\frac{\delta(\tau)\delta(\tau')  K(\Delta \tau)}{2\pi},
\end{equation}
with sgn$(x)$ being the signum function 
\begin{equation}
    \operatorname{sgn}(x)
    = \Theta(x)-\Theta(-x)
    =
    \begin{cases}
        1, & x>0,\\
        -1, & x<0,
    \end{cases}
\end{equation}
and where we introduced
\begin{equation}
    K(\Delta \tau) \equiv L_{0}(\Delta \tau) - L_{1}(\Delta \tau) = \log[\ii \Lambda(\Delta \tau - \ii \epsilon)].
    \label{eq:K_def}
\end{equation}
Now, we write (omitting the arguments of $A, B, ...$ for the sake of notational simplicity)
\begin{align}
(\partial_{\tau}\partial_{\tau'}W_{\tc{fw}}(\tau,\tau'))^2
    ={}&
    \frac{A^2}{(\Delta \tau-\ii\epsilon)^4}
    + \frac{2AB}{(\Delta \tau-\ii\epsilon)^3}
   \nonumber \\ &  + \frac{2AC+B^2}{(\Delta \tau-\ii\epsilon)^2}  
    + \frac{2BC}{\Delta \tau-\ii\epsilon}
    + C^2.
\label{eq:mathcalW00_schematic}
\end{align}
At this stage, Eq.~\eqref{eq:mathcalW00_schematic} cannot be interpreted as an identity of distributions without further qualification: the coefficients $A$, $B$, and $C$ contain discontinuous functions and Dirac deltas, and their products need not be defined in standard distribution theory. The formal expansion below is used only to identify the products for which an extension or regularization would be required; it does not itself specify such an extension. If one uses $\Theta(0)=1/2$ only in a formal sense, the coefficients in Eq.~\eqref{eq:mathcalW00_schematic} take the following expressions:
\begin{equation}
    A^2
    = \frac{1}{16\pi^2}
    \left[
    1 + 3\bigl(
    \Theta(\tau)\Theta(\tau')
    + \Theta(-\tau)\Theta(-\tau')
    \bigr)
    \right],
\end{equation}
\begin{equation}
    AB
    = \frac{3}{32\pi^2}
    \bigl(
    \delta(\tau')\operatorname{sgn}(\tau)
    - \delta(\tau)\operatorname{sgn}(\tau')
    \bigr),
\end{equation}
\begin{equation}
    AC
    = \frac{3}{16\pi^2}
    K(\Delta \tau)\delta(\tau)\delta(\tau'),
\end{equation}
\begin{equation}
    BC
    =\frac{\delta(\tau)\delta(\tau')     K(\Delta \tau)}{8\pi^2}
    \bigl(
\delta(\tau')\operatorname{sgn}(\tau)
    - \delta(\tau)\operatorname{sgn}(\tau')
    \bigr),
\end{equation}
and
\begin{equation}
    B^2
    = \frac{1}{16\pi^2}
    \bigl[
    \delta(\tau)^2 + \delta(\tau')^2
    - 2\delta(\tau)\delta(\tau')
    \operatorname{sgn}(\tau)\operatorname{sgn}(\tau')
    \bigr].
\end{equation}
The expressions above contain products such as $\Theta\delta$, $\delta\,\operatorname{sgn}$, and $\delta^2$ that are not canonically defined in standard distribution theory. In particular, assigning the point values $\Theta(0)=1/2$ and $\operatorname{sgn}(0)=0$ does not define these products. We do not select a regularization or extension prescription for them here. Consequently, the formal ``$\delta(0)$'' terms that appear below are diagnostic symbols indicating where such a prescription is required; they are not, by themselves, regulator-independent predictions for the detector response.

We shall carry out the analysis keeping track of each type of formal divergence. With this in mind, let us first focus on the terms in
Eq.~\eqref{eq:mathcalW00_schematic} that can produce contact divergences of the
form ``\(\delta(0)\)''. These are the last three terms, namely
\begin{equation}
    \mathcal{J}_2
    \equiv \int \dd \tau \, \dd \tau \,'\,
    \chi(\tau)\chi(\tau') e^{-\ii \Omega \Delta \tau}
    \frac{2AC+B^2}{(\Delta \tau-\ii\epsilon)^2},
    \label{eq:mathcalJ_2}
\end{equation}
\begin{equation}
    \mathcal{J}_1
    \equiv \int \dd \tau \, \dd \tau \,'\,
    \chi(\tau)\chi(\tau') e^{-\ii \Omega \Delta \tau}
    \frac{2BC}{\Delta \tau-\ii\epsilon},
    \label{eq:mathcalJ_1}
\end{equation}
and
\begin{equation}
    \mathcal{J}_0
    \equiv \int \dd \tau \, \dd \tau \,'\,
    \chi(\tau)\chi(\tau') e^{-\ii \Omega \Delta \tau}\, C^2.
    \label{eq:mathcalJ_0}
\end{equation}
Let us start by analyzing $\mathcal{J}_2$. As before, we introduce the
notation
\begin{equation}
    g(\tau,\tau') \equiv
    \chi(\tau)\chi(\tau') e^{-\ii \Omega \Delta \tau}.
    \label{eq:gdef}
\end{equation}
Explicitly, we have
\begin{align}
\mathcal{J}_2
={}& \frac{1}{16\pi^2}
\int_{-\infty}^{\infty}\dd\tau
\int_{-\infty}^{\infty}\dd\tau'\,
\frac{g(\tau,\tau')}{(\Delta\tau-\ii\epsilon)^2}
\,\mathcal{I}(\tau,\tau'),
\nonumber\\
\mathcal{I}(\tau,\tau')
\equiv{}&
6K(\Delta \tau)
\delta(\tau)\delta(\tau')
+ \delta(\tau)^2 + \delta(\tau')^2
\nonumber\\
&\quad
-2\,\delta(\tau)\delta(\tau')\,
\operatorname{sgn}(\tau)\operatorname{sgn}(\tau').
\end{align}
To handle this integral at the level of the formal expansion, we write
\begin{equation}
    \int_{-\infty}^{\infty}\dd x \,\, f(x)\,\delta(x)^2
    = \delta(0)\,f(0).
    \label{eq:formaldeltasquared}
\end{equation}
Moreover, the treatment of the $\delta\,\operatorname{sgn}$ term can be heuristically motivated by a symmetric smoothing in which $\delta$ is approximated by an even mollifier and $\operatorname{sgn}$ by an odd smooth function, both with the same vanishing width. Their product is odd and becomes localized at the origin. When paired with a smooth test function, the constant part of the test function cancels by parity, while the remaining contribution vanishes with the smoothing width. Accordingly, at the level of the formal expansion, we write
\begin{equation}
    \int_{-\infty}^{\infty}\dd x\,f(x)\delta(x)\operatorname{sgn}(x)
   =0.
    \label{eq:formal_delta_sgn}
\end{equation}
With these formal assignments,
\begin{align}
    \mathcal{J}_2
    ={}& \frac{1}{16\pi^2}\Biggl[
    \frac{
    6K(0)g(0,0)
    }{(-\ii\epsilon)^2}
        \nonumber\\
    & +  \delta(0)\int_{-\infty}^{\infty}\dd \tau \,'\,
    \frac{g(0,\tau')}{(\tau'+\ii\epsilon)^2}
    + \delta(0)\int_{-\infty}^{\infty}\dd \tau \,\,
    \frac{g(\tau,0)}{(\tau-\ii\epsilon)^2}
    \Biggr],
\end{align}
Using the definitions \eqref{eq:gdef} and \eqref{eq:K_def}
\begin{align}
    \mathcal{J}_2
    ={}& \frac{1}{16\pi^2}\Biggl[
    \frac{-6\log(\Lambda\epsilon)\chi(0)^2}{\epsilon^2}
 \nonumber \\ & + \chi(0)\delta(0)
    \int_{-\infty}^{\infty}\dd \tau \,\,
    \chi(\tau)
    \left(
    \frac{e^{\ii\Omega\tau}}{(\tau+\ii\epsilon)^2}
    + \frac{e^{-\ii\Omega\tau}}{(\tau-\ii\epsilon)^2}
    \right)\Biggr].
\end{align}
Let us now simplify $\mathcal{J}_1$
and $\mathcal{J}_0$ in order to obtain more information about the contact
divergences that may arise. For $\mathcal{J}_1$, we have
\begin{align}
    \mathcal{J}_1
   &  =\frac{1}{4\pi^2}
    \int_{-\infty}^{\infty}\dd \tau \,
    \int_{-\infty}^{\infty}\dd \tau \,'\,
    \frac{g(\tau,\tau') \delta(\tau)\delta(\tau')K(\Delta \tau)}{\Delta\tau-\ii\epsilon}
    \nonumber\\
    &\times
    \Bigl[
    \delta(\tau')\operatorname{sgn}(\tau)
    - \delta(\tau)\operatorname{sgn}(\tau')
    \Bigr].
\end{align}
Using the formal prescription of Eq.~\eqref{eq:formaldeltasquared}, it follows that
\begin{align}
    \mathcal{J}_1
    & =  \frac{\delta(0)}{4\pi^2}
    \int_{-\infty}^{\infty}\dd \tau \,\,
    \frac{
    g(\tau,0)K(\tau)
\delta(\tau)\operatorname{sgn}(\tau)}{\tau-\ii\epsilon}
    \nonumber\\
    &
    + \frac{\delta(0)}{4\pi^2}
    \int_{-\infty}^{\infty}\dd \tau \,'\,
    \frac{
    g(0,\tau')K(-\tau') \delta(\tau')\operatorname{sgn}(\tau')
    }{\tau'+\ii\epsilon}
.
\end{align}
To proceed with the evaluation, one could in principle use the prescription of Eq.~\eqref{eq:formal_delta_sgn}. Nonetheless, in the case of $\mathcal{J}_{1}$, this yields a formal term ``$\delta(0)\times 0$'', which
remains ambiguous. Let us therefore leave $\mathcal{J}_1$ in its present form
and proceed with the study of $\mathcal{J}_0$. Using the prescription of
Eq.~\eqref{eq:formaldeltasquared}, we can formally write
\begin{align}
    \mathcal{J}_0
    & = \frac{1}{4\pi^2}
    \int_{-\infty}^{\infty}\dd \tau \,
    \int_{-\infty}^{\infty}\dd \tau \,'\,
    g(\tau,\tau') (\delta(\tau)\delta(\tau'))^2
   K(\Delta \tau)^2
    \nonumber\\
    & = \frac{\chi(0)^2\log(\Lambda\epsilon)^2\,\delta(0)^2}{4\pi^2}.
\end{align}
The appearance of this formal expression indicates that the sharp-firewall expansion requires products of distributions for which no canonical definition is available. Hence, as formulated, the sharp-firewall construction does not canonically define the connected-correlation response function.

Let us now identify the origin of this obstruction. The first two terms in Eq.~\eqref{eq:mathcalW00_schematic}, when integrated against $g(\tau,\tau')$, involve the same kind of half-line distributional singularities that appear in the derivative-coupling response $\mathcal{F}^{(\pi)}_{\tc{fw}}$, discussed in Appendix~\ref{sec:appendix_distributions_firewall}. Their fate under a chosen extension prescription, including possible cancellations, is not analyzed here. The remaining terms in the formal expansion contain products such as $\delta(\tau)^2$ and $(\delta(\tau)\delta(\tau'))^2$, which do not possess a canonical definition in standard distribution theory. These products arise from differentiating the discontinuous wedge-indicator functions $f_{0}$, $f_{\tc{rf}}$, and $f_{\tc{fr}}$.

Physically, these functions encode the abrupt severing of correlations across the Rindler horizon. The calculation therefore suggests that the obstruction is tied to the discontinuous nature of the sharp-firewall idealization, rather than establishing a universal divergence for every possible realization of a firewall. For the sharp firewall, the quadratic response involves products of distributions that are not canonically defined.
Replacing the discontinuous wedge-indicator functions by a specified smooth transition profile of nonzero width would remove the delta distributions generated by their derivatives and hence eliminate these particular undefined products. This would, however, define a different regulated model.
Whether the finite-width response is well defined, and whether its sharp limit exists or is independent of the chosen smoothing, depend on the details of that construction and are left for future work.
Finite spatial detector smearing or a physical ultraviolet cutoff associated with a finite detector bandwidth could alter the singular structure. However, neither modification automatically removes the obstruction identified here. In the local non-linear coupling model, the noncanonical products arise at the level of the field correlation function, before it is integrated against the detector's spatial profile or frequency response. Multiplication by a smooth detector profile does not, by itself, define products such as $\delta^2$. Nevertheless, a detector model in which spatial smearing or finite bandwidth is incorporated before the composite field product is formed could modify this conclusion. Such models are not analyzed here.

The analogy between quadratic momentum coupling and energy-density coupling should be interpreted with care. The proportionality established in Sec.~\ref{sec:Minkowski_vacuum} holds for inertial pointlike detectors in the Minkowski vacuum and does not, by itself, extend to the firewall state. In particular, the energy-density correlation function contains both temporal and spatial derivative terms, and possible cancellations between these terms have not been analyzed here. We therefore do not claim that the energy-density response is established to be ill-defined in the sharp firewall state. Rather, the quadratic momentum calculation motivates the conjecture that similar difficulties may arise for local energy-density couplings.

\section{Conclusions}
\label{sec:conclusions}

In this work, we have investigated the response of Unruh-DeWitt detectors coupled to composite operators of a quantum scalar field, with particular emphasis on quadratic momentum coupling and the energy density. Our main objective was to understand the mathematical consistency of these detector models (to leading non-trivial order in perturbation theory) both in the Minkowski vacuum and in the Rindler firewall state.

We first revisited the response of pointlike detectors in the Minkowski vacuum. Rather than relying on integration by parts manipulations, we developed a distributional treatment based on the decomposition of singular kernels into Hadamard finite-part and Dirac-delta contributions. This approach yields compact expressions for the response functions of both derivative and quadratic momentum couplings, and extends straightforwardly to higher spacetime dimensions. In particular, we showed that, for inertial pointlike detectors in the vacuum, the response associated with the energy-density coupling is simply proportional to that of the quadratic momentum coupling, making the latter a convenient proxy for the former.

We then revisited the response of derivative-coupled detectors crossing the Rindler firewall. Using the same distributional framework, we reproduced the finite response originally obtained in Ref.~\cite{Louko2014}, making explicit the distributional cancellations that occur throughout the calculation. The resulting formalism provides a unified treatment of both vacuum and firewall response functions and naturally extends to more singular interaction models. We also found that the effect of the firewall on the detector response depends sensitively on the relative timing between the switching and the horizon crossing: depending on when the detector is switched on and off relative to the crossing, the firewall may either enhance or suppress the detector response with respect to the vacuum case.

The main result of this work concerns detector models coupled quadratically to the field momentum in the firewall state. Using normal ordering with respect to the firewall state as a formal diagnostic tool, we obtain a connected response function whose expansion contains products such as $\delta(\tau)^2$ and $(\delta(\tau)\delta(\tau'))^2$. Because these products have no canonical definition in standard distribution theory, the sharp, pointlike quadratic-momentum model does not determine a unique response without an additional extension or regularization prescription. This obstruction differs from the singularities encountered in the derivative-coupling model, where the complete response admits a well-defined distributional evaluation. In the present sharp, pointlike formulation, the noncanonical products arise directly from derivatives of the discontinuous wedge-indicator functions that encode the severing of correlations across the Rindler horizon. Based on our analysis in the Minkowski vacuum showing that the response of pointlike detectors coupled to the local energy density is proportional to that of detectors coupled to $\hat{\pi}^2$, this motivates the conjecture that the same distributional mechanism may also affect energy-density coupling in the sharp firewall model. Our results therefore identify an obstruction to defining a unique connected response function for quadratic-momentum coupling in the sharp, pointlike setting and suggest that related obstructions may arise for other composite local couplings.

These results suggest that the issues encountered here are not merely technical artifacts of the calculation, but rather signal a limitation of the sharp-firewall model with a pointlike detector and a quadratic local coupling. More precisely, in the sharp, pointlike formulation considered here, the connected response function is not uniquely determined at leading nontrivial perturbative order without an additional extension or regularization prescription. Moreover, the sharp Rindler construction does not determine the response to a finite-thickness black-hole firewall. An interesting direction for future work is the development of modified firewall models in which the severing of correlations occurs smoothly over a finite spacetime region. Such constructions may preserve the qualitative features of the firewall while removing the particular ill-defined products generated by derivatives of the discontinuous wedge-indicator functions. Finite spatial detector smearing or a physical ultraviolet cutoff associated with a finite detector bandwidth would not automatically resolve this obstruction, since the noncanonical products arise at the level of the field correlation function before integration against the detector profile. These modifications could nevertheless alter the singular structure and have not been analyzed here. It would also be interesting to investigate whether similar obstructions arise for other non-linear couplings and, more generally, to clarify the conditions under which detector models coupled to local observables remain mathematically well defined in non-Hadamard quantum states.

\acknowledgements

The authors thank Jorma Louko for the extremely interesting discussions and very helpful comments. MHZ thanks Prof. Achim Kempf for funding through his Dieter Schwarz grant. Research at Perimeter Institute is supported in part by the Government of Canada through the Department of Innovation, Science and Industry Canada and by the Province of Ontario through the Ministry of Colleges and Universities. EMM acknowledges support through the Discovery Grant Program of the Natural Sciences and Engineering Research Council of Canada (NSERC). EMM thanks the support from his Ontario Early Researcher award.

\onecolumngrid

\appendix

\section{Evaluation of the energy density correlation function in the vacuum state}
\label{sec:appendix_two_point}
In this Appendix, we explicitly show how to evaluate the two-point function $\langle 0 | \!:\!\hat{T}_{tt}(\mf x)\!: :\! \hat{T}_{tt}(\mf x')\!:\!| 0 \rangle$. We start by applying the renormalization via normal ordering, as in Eq. \eqref{eq:mean_value_zero}, so that
\begin{equation}
    \langle 0 | \!:\!\hat{T}_{tt}(\mf x)\!: :\! \hat{T}_{tt}(\mf x')\!:\!| 0 \rangle = \langle 0| \hat{T}_{tt}(\mf x)\hat{T}_{tt}(\mf x')| 0 \rangle - \langle 0| \hat{T}_{tt}(\mf x)| 0 \rangle \langle 0| \hat{T}_{tt}(\mf x')| 0 \rangle.
\end{equation}
Using
\begin{equation}
    \hat{T}_{tt}(\mf x) = \frac{1}{2}((\partial_{t}\hat{\phi}(\mf x))^2 + \partial_{i}\hat{\phi}(\mf x)\partial^{i}\hat{\phi}(\mf x)),
\end{equation}
together with the auxiliary definitions $\hat{\psi}(\mf x) = \partial_{t}\hat{\phi}(\mf x)$ and $\hat{\varphi}_i (\mf x) = \partial_i \hat{\phi}(\mf x)$, we have
\begin{equation}
  \langle 0| \hat{T}_{tt}(\mf x)\hat{T}_{tt}(\mf x')| 0 \rangle = \frac{1}{4}\left[\langle 0 | \hat{\psi}^2(\mf x) \hat{\psi}^2(\mf x')| 0 \rangle +   \sum_{i}\langle 0 | \hat{\psi}^2(\mf x) \hat{\varphi}_{i}^2(\mf x')| 0 \rangle + \sum_{i}\langle 0 | \hat{\varphi}_{i}^2(\mf x) \hat{\psi}^2(\mf x')| 0 \rangle + \sum_{i, j}\langle 0 | \hat{\varphi}_{i}^2(\mf x) \hat{\varphi}_{j}^2(\mf x')| 0 \rangle \right]. 
\end{equation}
Each one of the individual terms in the equation above can be simplified by using Wick's theorem, or a similar procedure as employed in Appendix A of reference \cite{Sachs1}, we can write
\begin{equation}
    \langle 0 | \hat{\psi}^2(\mf x) \hat{\psi}^2(\mf x')| 0 \rangle = 2 (\langle 0| \hat{\psi}(\mf x)\hat{\psi}(\mf x')|0 \rangle)^2 + \langle 0| \hat{\psi}^2(\mf x)|0\rangle \langle 0| \hat{\psi}^2(\mf x')|0\rangle,
\end{equation}
\begin{equation}
        \langle 0 | \hat{\psi}^2(\mf x) \hat{\varphi}_{i}^2(\mf x')| 0 \rangle = 2 (\langle 0| \hat{\psi}(\mf x)\hat{\varphi}_{i}(\mf x')|0 \rangle)^2 + \langle 0| \hat{\psi}^2(\mf x)|0\rangle \langle 0| \hat{\varphi}_{i}^2(\mf x')|0\rangle,
        \label{eq:use_for_momentum}
\end{equation}
\begin{equation}
        \langle 0 | \hat{\varphi}_{i}^2(\mf x) \hat{\psi}^2(\mf x')| 0 \rangle = 2 (\langle 0| \hat{\varphi}_{i}(\mf x)\hat{\psi}(\mf x')|0 \rangle)^2 + \langle 0| \hat{\varphi}_{i}^2(\mf x)|0\rangle \langle 0| \hat{\psi}^2(\mf x')|0\rangle,
\end{equation}
and
\begin{equation}
            \langle 0 | \hat{\varphi}_{i}^2(\mf x) \hat{\varphi}_{j}^2(\mf x')| 0 \rangle = 2 (\langle 0| \hat{\varphi}_{i}(\mf x)\hat{\varphi}_{j}(\mf x')|0 \rangle)^2 + \langle 0| \hat{\varphi}_{i}^2(\mf x)|0\rangle \langle 0| \hat{\varphi}_{j}^2(\mf x')|0\rangle.
\end{equation}
On the other hand,
\begin{align}
    \langle 0|\hat{T}_{tt}(\mf x)|0\rangle\,\langle 0|\hat{T}_{tt}(\mf x')|0\rangle
    &= \frac{1}{4}\Bigl[
        \langle 0|\hat{\psi}^2(\mf x)|0\rangle\,\langle 0|\hat{\psi}^2(\mf x')|0\rangle
        + \sum_{i}\langle 0|\hat{\psi}^2(\mf x)|0\rangle\,\langle 0|\hat{\varphi}_{i}^2(\mf x')|0\rangle \nonumber\\
    &\quad{} + \sum_{i}\langle 0|\hat{\varphi}_{i}^2(\mf x)|0\rangle\,\langle 0|\hat{\psi}^2(\mf x')|0\rangle
        + \sum_{i,j}\langle 0|\hat{\varphi}_{i}^2(\mf x)|0\rangle\,\langle 0|\hat{\varphi}_{j}^2(\mf x')|0\rangle
    \Bigr].
\end{align}
Therefore,
\begin{align}
   \langle 0 | \!:\!\hat{T}_{tt}(\mf x)\!: :\! \hat{T}_{tt}(\mf x')\!:\!| 0 \rangle  &=\frac{1}{2} \Biggl[(\langle 0| \hat{\psi}(\mf x)\hat{\psi}(\mf x')|0 \rangle)^2 + \sum_{i}(\langle 0| \hat{\psi}(\mf x)\hat{\varphi}_{i}(\mf x')|0 \rangle)^2 + 
    \sum_{i}(\langle 0| \hat{\varphi}_{i}(\mf x)\hat{\psi}(\mf x')|0 \rangle)^2 \nonumber \\ & \qquad + \sum_{i,j}(\langle 0| \hat{\varphi}_{i}(\mf x)\hat{\varphi}_{j}(\mf x')|0 \rangle)^2 \Biggr].
\end{align}
In terms of the Minkowski vacuum Wightman function $W_{0}(\mf x, \mf x)$, we have
\begin{equation}
    \langle 0 | \!:\!\hat{T}_{tt}(\mf x)\!: :\! \hat{T}_{tt}(\mf x')\!:\!| 0 \rangle  =\frac{1}{2} \left[(\partial_{t}\partial_{t'}W_{0}(\mf x, \mf x'))^2 +  \sum_{i}(\partial_{i}\partial_{t'}W_{0}(\mf x, \mf x'))^2 + \sum_{i}(\partial_{t}\partial_{i'}W_{0}(\mf x, \mf x'))^2 + \sum_{i, j}(\partial_{i}\partial_{j'}W_{0}(\mf x, \mf x'))^2\right].
\end{equation}
Those calculations can be easily adapted to evaluate the two-point function for the quadratic momentum coupling. Indeed, the starting point is
\begin{equation}
    \langle 0 |\mathopen{:}\hat{\pi}^2(\mf x)\mathopen{:} \mathopen{:}\hat{\pi}^2(\mf x')\mathopen{:} | 0 \rangle
    =
    \langle 0 | \hat{\pi}^2(\mf x)\hat{\pi}^2(\mf x') |0\rangle
    -
    \langle 0 | \hat{\pi}^2(\mf x)|0\rangle
    \langle 0 | \hat{\pi}^2(\mf x')|0\rangle .
\end{equation}
Since $\hat{\pi} = \partial_{t}\hat{\phi}$, we can use Eq.~\eqref{eq:use_for_momentum} to obtain
\begin{equation}
    \langle 0 |\mathopen{:}\hat{\pi}^2(\mf x)\mathopen{:} \mathopen{:}\hat{\pi}^2(\mf x')\mathopen{:} | 0 \rangle = 2 (\langle 0|\partial_{t}\hat{\phi}(\mf x)\partial_{t'}\hat{\phi}(\mf x') |0\rangle)^2 = 2 (\partial_{t}\partial_{t'}W_{0}(\mf x, \mf x'))^2.
\end{equation}

\section{Important distributional identities}
\label{sec:Vanilla_calculations}

In this appendix, we derive important distributional identities that are used in the main text. These results are included here for the sake of completeness and for establishing notation. The interested reader can find more details in \cite{GelfandShilov2016}. We start by considering integrals of the form
\begin{equation}
    I_{\epsilon}[f] = \int_{-\infty}^{\infty}\dd x \,\frac{ f(x)}{(x \pm \ii \epsilon)^n}.
\end{equation}
Assuming that $f(x)$ is smooth and that $\lim_{x\to \pm \infty}f(x) = 0$, we can apply integration by parts $(n - 1)$ times:
\begin{equation}
    I_{\epsilon}[f] =\frac{1}{(n - 1)!}\int_{-\infty}^{\infty}\dd x \,\frac{f^{(n - 1)}(x)}{(x \pm \ii \epsilon)}.
\end{equation}
Now, this integral can be solved using the Sokhotski–Plemelj theorem. If $h$ is any complex-valued functions that is defined and continuous on $(a, b)$, with $a < 0 < b$, then
\begin{equation}
    \lim_{\epsilon \to 0^{+}} \int_{a}^{b}\dd x \,\frac{h(x)}{x \pm \ii \epsilon} = \mp \ii \pi h(0) + \operatorname{PV} \int_{a}^{b} \dd x \, \frac{h(x)}{x},
    \label{eq:sokkkkjkjkl}
\end{equation}
where $\operatorname{PV}$ indicates that we are evaluating the Cauchy principal value, which can be defined as
\begin{equation}
    \operatorname{PV} \int_{a}^{b}\dd x \, \frac{h(x)}{x} = \lim_{\delta\to 0^{+}}\left[\int_{a}^{-\delta}  + \int_{\delta}^{b}\right]\dd x \,\frac{h(x)}{x}.
\end{equation}
Therefore,
\begin{equation}
    \lim_{\epsilon \to 0^{+}}I_{\epsilon}[f] = \frac{1}{(n - 1)!}\left[\mp \ii \pi f^{(n - 1)}(0)  + \operatorname{PV}\int_{-\infty}^{\infty}{\dd x \, \frac{f^{(n - 1)}(x)}{x}}\right].
    \label{eq:Iepsilon}
\end{equation}
When evaluating the Cauchy Principal value over the entire real line, it is often convenient to use the following formula (especially for numerical evaluation purposes)
\begin{equation}
    \operatorname{PV}\int_{-\infty}^{\infty}\dd x \, \frac{h(x)}{x} = \lim_{r \to 0^{+}}\int_{r}^{\infty}\dd x \,\frac{h(x) - h(-x)}{x}.
\end{equation}
There is yet another way to obtain the formula in Eq.~\eqref{eq:Iepsilon}. Instead of applying integration by parts, one can work directly with the distribution
\begin{equation}
\lim_{\epsilon \to 0^{+}} \frac{1}{(x -\ii \epsilon)^n} \equiv \frac{1}{(x -\ii 0^{+})^n}.
\label{eq:fepsilondistribution}
\end{equation}
In this derivation, the starting point is the Sokhotski–Plemelj theorem for distributions \cite{GelfandShilov2016}:
\begin{equation}\frac{1}{x \mp \ii 0^{+}}=\operatorname{PV}\left(\frac{1}{x}\right) \pm \ii \pi \delta(x).
    \label{eq:SokhotskiPlemeljdist}
\end{equation}
The goal is, then, to generalize this formula for a general $n \in \mathbb{N}$, as in Eq.~\eqref{eq:fepsilondistribution}. To accomplish this,  we need to take derivatives of the distributions in Eq.~\eqref{eq:SokhotskiPlemeljdist}.
Let $\langle \mathcal{D}, f \rangle$ denote the action of the distribution $\mathcal{D}$ on the test function $f$. Then, the $k$-th derivative of the distribution $\mathcal{D}$ is defined as
\begin{equation}
    \langle \mathcal{D}^{(k)}, f \rangle = (-1)^{k}  \langle \mathcal{D}, f^{(k)} \rangle  
\end{equation}
In our case, one can evaluate
\begin{equation}
    \frac{d^k}{d x^k} \frac{1}{x \mp \ii 0^{+}} = (-1)^{k}k! \frac{1}{(x \mp \ii 0^{+})^{(k+1)}},
\end{equation}
whereas $\delta^{(k)}(x)$ acts on a test function $f$ as
\begin{equation}
   \langle \delta^{(k)}, f \rangle = (-1)^{k}f^{(k)}(0).
\end{equation}
To handle the derivative of the Principal Value distribution, we can use the Hadamard Finite Part (FP) distribution. Let $P_{q}(x)$ denote the Taylor polynomial of the test function $f$ centered at $x = 0$, namely
\begin{equation}
    P_{q}(x) = \sum_{n = 0}^{q}\frac{f^{(n)}(0)}{n!}x^{n}.
\end{equation}
Define
\begin{equation}
    f_{q}(x) \equiv f(x) - P_{q}(x).
\end{equation}
Then, we can write \cite{GelfandShilov2016}
\begin{equation}
  \left\langle \operatorname{FP}\!\left(\frac{1}{x^{n}}\right), f \right\rangle
  = \lim_{r \to 0^{+}} \int_{r}^{\infty} \dd x \,\, \frac{f_{n-1}(x) + (-1)^{n} f_{n-1}(-x)}{x^{n}} .
\end{equation}
Now, it turns out that the distributions $\operatorname{PV}$ and $\operatorname{FP}$ are related via
\begin{equation}
\left\langle \operatorname{FP}\!\left(\frac{1}{(x - x_0)^{n}}\right), f \right\rangle = \frac{1}{(n - 1)!} \left\langle \operatorname{PV}\left(\frac{1}{x - x_0}\right), f^{(n - 1)}\right\rangle.
\label{eq:PV_and_FP}
\end{equation}
This result can be established by induction (see also \cite{BarataNotasFisicaMatematica}). Indeed, assume Eq.~\eqref{eq:PV_and_FP} holds for $n = k - 1 \in \mathbb{N}$. Take $x_{0} = 0$ without loss of generality. For $n = k$, we have (for any $r > 0$)
\begin{equation}
    \int_{-\infty}^{-r} \dd x \, \frac{f(x)}{x^k} = \frac{1}{1 - k}f(x)x^{1 - k}\Bigg|^{-r}_{-\infty} + \frac{1}{k - 1}\int_{-\infty}^{-r}\dd x \, \frac{f'(x)}{x^{k -1}},
\end{equation}
and
\begin{equation}
    \int_{r}^{\infty} \dd x \, \frac{f(x)}{x^k} = \frac{1}{1 - k}f(x)x^{1 - k}\Bigg|^{\infty}_{r} + \frac{1}{k - 1}\int_{-r}^{\infty}\dd x \, \frac{f'(x)}{x^{k -1}}.
\end{equation}
Therefore,
\begin{equation}
    \int_{-\infty}^{-r} \dd x \, \frac{f(x)}{x^k} + \int_{r}^{\infty} \dd x \, \frac{f(x)}{x^k} = \frac{1}{k -1} \left[\frac{f(r)}{r^{k - 1}} - (-1)^{k - 1}\frac{f(-r)}{r^{k - 1}} +\int_{-\infty}^{-r}\dd x \, \frac{f'(x)}{x^{k -1}} +  \int_{-r}^{\infty}\dd x \, \frac{f'(x)}{x^{k -1}}\right].
    \label{eq:IBP12232}
\end{equation}
Now, let us write
\begin{equation}
    \int_{-r}^{r}\dd x \, \frac{f(x)}{x^k} = F_{1}(r) + D_{1}(r),
    \label{eq:somestep}
\end{equation}
where $ F_{1}(r)$ is the finite part of the integral, and  $D_{1}(r)$ is a function that diverges on the limit $r \to 0$. Similarly, we have
\begin{equation}
    \int_{-r}^{r}\dd x \, \frac{f'(x)}{x^{k -1}} = F_{2}(r) + D_{2}(r),
    \label{eq:somestep2}
\end{equation}
By Eq.~\eqref{eq:IBP12232}, it follows that
\begin{equation}
    D_{1}(r) =  D_{2}(r) + \frac{1}{k -1} \left[\frac{f(r)}{r^{k - 1}} - (-1)^{k - 1}\frac{f(-r)}{r^{k - 1}}\right],
\end{equation}
so that the finite part of the integrals in Eqs. \eqref{eq:somestep} and \eqref{eq:somestep2} are the same, i.e., $F_{1}(r) = F_{2}(r)$. Thus, 
\begin{equation}
    \operatorname{FP}\int_{-\infty}^{\infty}\dd x \, \frac{f(x)}{x^k} = \frac{1}{k - 1}\operatorname{FP}\int_{-\infty}^{\infty}\dd x \, \frac{f'(x)}{x^{k - 1}} = \frac{1}{(k-1)!}\operatorname{PV}\int_{-\infty}^{\infty}\dd x \, \frac{f^{k - 1}}{x},
\end{equation}
by the induction hypothesis. Therefore, the result of Eq.~\eqref{eq:PV_and_FP} holds. Then, using this result, the following distributional identity can be established:
\begin{equation}
    \frac{1}{(x \mp \ii 0^{+})^n} = \operatorname{FP}\left(\frac{1}{x^n} \right) \pm \frac{(-1)^{n - 1}}{(n - 1)!} \ii \pi \delta^{(n - 1)}(x),
    \label{eq:generalizedSP}
\end{equation}
Therefore, for any $n \in \mathbb{N}$ and any function $f \in \mathcal{C}^{n - 1}(\mathbb{R})$, we can use \eqref{eq:generalizedSP} and \eqref{eq:PV_and_FP} to obtain the same result as before:
\begin{align}
 \lim_{\epsilon \to 0^{+}}I_{\epsilon}[f]  & = \mp\frac{\ii \pi}{(n - 1)!} f^{(n - 1)}(0) + \operatorname{FP}\int_{-\infty}^{\infty}\dd x \, \frac{f(x)}{x^n} \nonumber \\
   & = \frac{1}{(n - 1)!}\left(\mp\ii \pi f^{(n - 1)}(0) + \operatorname{PV}\int_{-\infty}^{\infty}\dd x \, \frac{f^{(n - 1)}(x)}{x}\right).
    \label{eq:that_is_the_one_appendix}
\end{align}

\section{Derivation of firewall response function in the derivative-coupling model using distributions}
\label{sec:appendix_distributions_firewall}

In this appendix, we show all the calculations necessary to obtain Eq.~\eqref{eq:requiem_main_text} using the full expression of the Wightman function in the firewall state,
\begin{equation}
     W_{\tc{fw}}(\tau,  \tau')
     = -\frac{1}{4\pi}\log[-\Lambda^2(\Delta \tau - \ii \epsilon)^2]\; f_{0}(\tau, \tau')
  - \frac{1}{4\pi}\log[\Lambda(\epsilon + \ii \Delta \tau)]\;\bigl(f_{\tc{rf}}(\tau, \tau') + f_{\tc{fr}}(\tau, \tau')\bigr),
\end{equation}
along with a careful evaluation of the distributional limits $\epsilon \to 0^{+}$. Evaluating the derivatives of this expression, it follows that
\begin{align}
\partial_{\tau}\partial_{\tau'}W_{\tc{fw}}(\tau,\tau')
= & -\frac{1}{4\pi}\Biggl[\frac{2f_{0}(\tau, \tau') + f_{\tc{rf}}(\tau, \tau') + f_{\tc{fr}}(\tau, \tau')}{(\Delta \tau - \ii \epsilon)^2}  + 2\frac{\partial_{\tau'}f_{0}(\tau, \tau') - \partial_{\tau}f_{0}(\tau, \tau')}{\Delta \tau - \ii \epsilon} \nonumber \\ & + \frac{\partial_{\tau'}f_{\tc{rf}}(\tau, \tau') + \partial_{\tau'}f_{\tc{fr}}(\tau, \tau') - \partial_{\tau}f_{\tc{rf}}(\tau, \tau') - \partial_{\tau}f_{\tc{fr}}(\tau, \tau')}{\Delta \tau - \ii \epsilon} \nonumber \\ & + L_{0}(\Delta \tau) \partial_{\tau}\partial_{\tau'}f_{0}(\tau, \tau') +  L_{1}(\Delta \tau)( \partial_{\tau}\partial_{\tau'}f_{\tc{rf}}(\tau, \tau')+ \partial_{\tau}\partial_{\tau'}f_{\tc{fr}}(\tau, \tau'))\Biggr].
\label{eq:dtaudtaup_Wfw_general}
\end{align}
Thus, we need to evaluate (with $g(\tau, \tau') \equiv \chi(\tau)\chi(\tau')e^{-\ii \Omega \Delta \tau}$ )
\begin{align}
    \mathcal{F}^{(\pi)}_{\tc{fw}}  & = -\frac{1}{4\pi}\int_{-\infty}^{\infty}\dd \tau \, \int_{-\infty}^{\infty}\dd \tau \,' g(\tau, \tau') \Biggl[\frac{2f_{0}(\tau, \tau') + f_{\tc{rf}}(\tau, \tau') + f_{\tc{fr}}(\tau, \tau')}{(\Delta \tau - \ii \epsilon)^2}  \nonumber \\ & + 2\frac{\partial_{\tau'}f_{0}(\tau, \tau') - \partial_{\tau}f_{0}(\tau, \tau')}{\Delta \tau - \ii \epsilon} \nonumber \\ & + \frac{\partial_{\tau'}f_{\tc{rf}}(\tau, \tau') + \partial_{\tau'}f_{\tc{fr}}(\tau, \tau') - \partial_{\tau}f_{\tc{rf}}(\tau, \tau') - \partial_{\tau}f_{\tc{fr}}(\tau, \tau')}{\Delta \tau - \ii \epsilon} \nonumber \\ & + L_{0}(\Delta \tau) \partial_{\tau}\partial_{\tau'}f_{0}(\tau, \tau') +  L_{1}(\Delta \tau)( \partial_{\tau}\partial_{\tau'}f_{\tc{rf}}(\tau, \tau')+ \partial_{\tau}\partial_{\tau'}f_{\tc{fr}}(\tau, \tau'))\Biggr].
    \label{eq:nemesis}
\end{align}
For the sake of organization, let us define
\begin{equation}
    \mathcal{A} \equiv \int_{-\infty}^{\infty}\int_{-\infty}^{\infty} \dd \tau \, \dd \tau \,' \frac{g(\tau, \tau')}{(\Delta \tau - \ii \epsilon)^2}\Biggl[f_{0}(\tau, \tau') + \frac{f_{\tc{rf}}(\tau, \tau') + f_{\tc{fr}}(\tau, \tau')}{2} \Biggr],
\end{equation}
\begin{align}
\mathcal{B} \equiv  &- \int_{-\infty}^{\infty}\int_{-\infty}^{\infty} \dd \tau \, \dd \tau \,' \frac{g(\tau, \tau')}{\Delta \tau - \ii \epsilon}\Biggl[\partial_{\tau}f_{0}(\tau, \tau') - \partial_{\tau'}f_{0}(\tau, \tau') \nonumber \\ & + \frac{\partial_{\tau}f_{\tc{rf}}(\tau, \tau') + \partial_{\tau}f_{\tc{fr}}(\tau, \tau') - \partial_{\tau'}f_{\tc{rf}}(\tau, \tau') - \partial_{\tau'}f_{\tc{fr}}(\tau, \tau')}{2}\Biggr],
\label{eq:term_mathcal_B_expanded}
\end{align}
and
\begin{align}
    \mathcal{C} \equiv \frac{1}{2}\int_{-\infty}^{\infty}\int_{-\infty}^{\infty} \dd \tau \, \dd \tau \,' g(\tau, \tau')&\Biggl[ L_{0}(\Delta \tau) \partial_{\tau}\partial_{\tau'}f_{0}(\tau, \tau') +  L_{1}(\Delta \tau) \partial_{\tau}\partial_{\tau'}f_{\tc{rf}}(\tau, \tau') \nonumber \\ & +  L_{1}(\Delta \tau) \partial_{\tau}\partial_{\tau'}f_{\tc{fr}}(\tau, \tau')\Biggr].
    \label{eq:term_mathcal_C_expanded}
\end{align}
This way, Eq.~\eqref{eq:nemesis} can be compactly written as
\begin{equation}
    \mathcal{F}^{(\pi)}_{\tc{fw}} = -\frac{1}{2\pi}(\mathcal{A} + \mathcal{B} + \mathcal{C}).
\end{equation}
The distributional derivatives of the auxiliary functions are (assuming $x_{0} > 0$)
\begin{align}
    \partial_{\tau}f_{0} & = -\delta(\tau)\Theta(-\tau')\Theta(\tau + 2x_{0})\Theta(\tau' + 2x_{0}) + \delta(\tau + 2x_{0})\Theta(-\tau)\Theta(-\tau')\Theta(\tau' + 2x_{0}) + \delta(\tau)\Theta(\tau') \nonumber \\ & =
    \Theta(-\tau')\Theta(\tau' + 2x_{0})(\delta(\tau + 2x_{0}) - \delta(\tau)) + \delta(\tau)\Theta(\tau')
    \label{eq:partial_tau_f0}
\end{align}
\begin{align}
    \partial_{\tau'}f_{0} & = -\delta(\tau')\Theta(-\tau)\Theta(\tau + 2x_{0})\Theta(\tau' + 2x_{0}) + \Theta(-\tau)\Theta(-\tau')\Theta(\tau + 2x_{0})\delta(\tau' + 2x_{0}) + \delta(\tau')\Theta(\tau) \nonumber \\ 
    & = \Theta(-\tau)\Theta(\tau + 2x_{0})(\delta(\tau' + 2x_{0}) - \delta(\tau')) + \delta(\tau')\Theta(\tau)
    \label{eq:partial_tauprime_f0}
\end{align}
\begin{equation}
    \partial_{\tau}\partial_{\tau'}f_{0} = 2\delta(\tau)\delta(\tau') - \delta(\tau + 2x_{0})\delta(\tau') - \delta(\tau' + 2x_{0})\delta(\tau) + \delta(\tau + 2x_{0})\delta(\tau' + 2x_{0})
\end{equation}
\begin{equation}
    \partial_{\tau}f_{\tc{rf}} 
    = \Theta(\tau')(\delta(\tau + 2x_{0}) - \delta(\tau))
    \label{eq:partial_taurf}
\end{equation}
\begin{equation}
    \partial_{\tau'}f_{\tc{rf}} 
    = \delta(\tau')\Theta(-\tau)\Theta(\tau + 2x_{0})
    \label{eq:partial_tauprimerf}
\end{equation}
\begin{equation}
    \partial_{\tau}\partial_{\tau'}f_{\tc{rf}}
    = \delta(\tau')(\delta(\tau + 2x_{0}) - \delta(\tau))
\end{equation}
\begin{equation}
    \partial_{\tau}f_{\tc{fr}} 
    = \delta(\tau)\Theta(-\tau')\Theta(\tau' + 2x_{0})
    \label{eq:partial_taufr}
\end{equation}
\begin{equation}
    \partial_{\tau'}f_{\tc{fr}} 
    = \Theta(\tau)(\delta(\tau' + 2x_{0}) - \delta(\tau'))
    \label{eq:partial_tauprimefr}
\end{equation}
\begin{equation}
 \partial_{\tau}\partial_{\tau'}f_{\tc{fr}}
    = \delta(\tau)(\delta(\tau' + 2x_{0}) - \delta(\tau'))
\end{equation}
Let us start dealing with $\mathcal{A}$. 
Using the definitions expressed in Eq.~\eqref{eq:f_auxiliary_tau_tau'}, we have
\begin{align}
    \mathcal{A} = &  \Biggl[\int_{-2a}^{0}\dd \tau \, \int_{-2a}^{0}\dd \tau \,' + \int_{0}^{\infty}\dd \tau \, \int_{0}^{\infty}\dd \tau \,'  \Biggr]\frac{\chi(\tau)\chi(\tau')e^{-\ii \Omega \Delta \tau}}{(\Delta \tau - \ii \epsilon)^2} \nonumber \\ & + \frac{1}{2}\int_{-2a}^{0}\dd \tau \, \int_{0}^{\infty} \dd \tau \,' \frac{\chi(\tau)\chi(\tau')e^{-\ii \Omega \Delta \tau}}{(\Delta \tau - \ii \epsilon)^2} + \frac{1}{2}\int_{0}^{\infty}\dd \tau \, \int_{-2a}^{0} \dd \tau \,' \frac{\chi(\tau)\chi(\tau')e^{-\ii \Omega \Delta \tau}}{(\Delta \tau - \ii \epsilon)^2}.
\label{eq:term_mathcal_A_expanded}
\end{align}
At this point, we can use assumptions about $\chi(\tau)$ to simplify this expression. Given our physical setup, it is natural to assume that $\chi(\tau)$ is compactly supported. More than that, since the detector will only cross the firewall between the edges R and F, we have to assume that $\chi(\tau) = 0$ for $\tau < -2a$. Thus, we can write (where we recall the definition $g(\tau, \tau') \equiv \chi(\tau)\chi(\tau')e^{-\ii \Omega \Delta \tau}$)
\begin{align}
   \mathcal{A} = &  \Biggl[\int_{-\infty}^{0}\dd \tau \, \int_{-\infty}^{0}\dd \tau \,' + \int_{0}^{\infty}\dd \tau \, \int_{0}^{\infty}\dd \tau \,'  \Biggr]\frac{g(\tau, \tau')}{(\Delta \tau - \ii \epsilon)^2} \nonumber \\ & + \frac{1}{2}\int_{-\infty}^{0}\dd \tau \, \int_{0}^{\infty} \dd \tau \,' \frac{g(\tau, \tau')}{(\Delta \tau - \ii \epsilon)^2} + \frac{1}{2}\int_{0}^{\infty}\dd \tau \, \int_{-\infty}^{0} \dd \tau \,' \frac{g(\tau, \tau')}{(\Delta \tau - \ii \epsilon)^2}.
\end{align}
Before simplifying this expression and evaluating the limit $\epsilon \to 0^{+}$ with the aid of distributions, it turns out that this term can be directly related to the detector's response function when the field is prepared in the {\it Minkowski vacuum}, $\mathcal{F}^{(\pi)}_{0}$. Indeed, it is straightforward to show that
\begin{equation}
     \mathcal{F}^{(\pi)}_{0} = -\frac{1}{2 \pi} \int_{-\infty}^{\infty}\int_{-\infty}^{\infty} \dd \tau \, \dd \tau \,'\chi(\tau)\chi(\tau')\frac{e^{-\ii \Omega \Delta \tau}}{(\Delta \tau -\ii \epsilon)^2} \equiv -\frac{1}{2\pi}\mathcal{A}_{0}.
\end{equation}
It turns out that the term $\mathcal{A}_{0}$ can be compactly written as 
\begin{equation}
    \mathcal{A}_{0} = \int_{-\infty}^{\infty}\dd u \, \frac{\Phi(u)}{(u - \ii \epsilon)^2},
\end{equation}
where $\Phi(u) = e^{-\ii \Omega u}F(u)$, with
\begin{equation}
    F(u) =  \int_{-\infty}^{\infty}{\dd s \chi(s)\chi(s - u)}.
\end{equation}
Let us now show how to evaluate $\mathcal{A}_{0}$, and thus $\mathcal{F}^{(\pi)}_{0}$, in detail. Recall the fundamental identity in the theory of distributions,
\begin{equation}
    \lim_{\epsilon \to 0^{+}}\frac{1}{(x \mp \ii \epsilon)^n} = \operatorname{FP}\left(\frac{1}{x^n} \right) \pm \frac{(-1)^{n - 1}}{(n - 1)!} \ii \pi \delta^{(n - 1)}(x).
    \label{eq:generalizedSP2}
\end{equation}
Taking $n=2$, this expression reads
\begin{equation}
      \lim_{\epsilon \to 0^{+}}\frac{1}{(x \mp \ii \epsilon)^2} = \operatorname{FP}\left(\frac{1}{x^2} \right) \mp \ii \pi \delta'(x).
\end{equation}
Thus, it follows that
\begin{equation}
    \mathcal{A}_{0} = \int_{-\infty}^{\infty}\dd u \, \frac{\Phi(u)}{(u - \ii \epsilon)^2} = 
    \ii \pi \Phi'(0) + \operatorname{FP}\int_{-\infty}^{\infty}\dd u \, \frac{\Phi(u)}{u^2}.
\end{equation}
Assuming the switching function $\chi$ to be such that $\lim_{\tau \pm \infty}\chi(\tau) = 0$, we get $\Phi'(0) = -\ii \Omega F(0)$. Then,
\begin{align}
    \mathcal{A}_{0} & = \pi \Omega F(0) + \int_{0}^{\infty}\dd u \, \frac{\Phi(u) + \Phi(-u) - 2\Phi(0)}{u^2} \nonumber \\
    & =\pi \Omega F(0) + 2\int_{0}^{\infty}\dd u \, \frac{\cos(\Omega u)F(u) - F(0)}{u^2}. 
\end{align}
Therefore,
\begin{equation}
    \mathcal{F}^{(\pi)}_{0}  = -\frac{\Omega}{2}F(0) + \frac{1}{\pi}\int_{0}^{\infty}\dd u \, \frac{F(0) - F(u)\cos(\Omega u)}{u^2}.
\end{equation}
Let us now show that this formula matches Eq. (3.3b) in Ref.~\cite{Louko2014}. Indeed, 
\begin{align}
\mathcal{F}^{(\pi)}_{0} & = -\frac{\Omega}{2}F(0) + \frac{1}{\pi}\int_{0}^{\infty}\dd u \, \frac{F(0) - \cos(\Omega u)F(u)}{u^2} \nonumber \\ & = -\frac{\Omega}{2}F(0) + \frac{1}{\pi}\int_{0}^{\infty}\dd u \, \frac{F(0) - \cos(\Omega u)F(u) + F(0)\cos(\Omega u) - F(0)\cos(\Omega u )}{u^2} \nonumber \\ & = -\frac{\Omega}{2}F(0) + \frac{1}{\pi}\int_{0}^{\infty}\dd u \, \frac{\cos(\Omega u)(F(0) - F(u))}{u^2} - \frac{F(0)}{\pi}\int_{0}^{\infty}\dd u \, \frac{\cos(\Omega u) - 1}{u^2} \nonumber \\
     &= \frac{1}{\pi}\int_{0}^{\infty}\dd u \, \frac{\cos(\Omega u)(F(0) - F(u))}{u^2} + \frac{|\Omega|}{2}F(0) - \frac{\Omega}{2}F(0).
\end{align}
Finally, using
\begin{equation}
    \frac{1}{2}\left(|\Omega| - \Omega\right) = -\Omega \Theta(-\Omega),
\end{equation}
and observing that
\begin{equation}
    F(0) = \int_{-\infty}^{\infty}\dd s \chi(s)^2,
\end{equation}
we can cast the Minkowski response function into
\begin{equation}
    \mathcal{F}^{(\pi)}_{0} = -\Omega \Theta(-\Omega)\int_{-\infty}^{\infty}\dd s \chi(s)^2 + \frac{1}{\pi}\int_{0}^{\infty}\dd u \, \frac{\cos(\Omega u)}{u^2}\int_{-\infty}^{\infty}\dd s \chi(s)(\chi(s) - \chi(s-u)),
\end{equation}
which is exactly the result of Eq. (3.3b) in reference \cite{Louko2014}. 

Next, let us use $\mathcal{A}_{0}$ to simplify $\mathcal{A}$. First, notice that
\begin{equation}
    \mathcal{A}_{0} = \Biggl(\int_{-\infty}^{0}\dd \tau \, \int_{-\infty}^{0}\dd \tau \,' + \int_{0}^{\infty}\dd \tau \, \int_{0}^{\infty}\dd \tau \,' + \int_{0}^{\infty}\dd \tau \, \int_{-\infty}^{0}\dd \tau \,' + \int_{-\infty}^{0}\dd \tau \,' \int_{0}^{\infty}\dd \tau \, \Biggr)\frac{g(\tau, \tau')}{(\Delta \tau - \ii \epsilon)^2}
\end{equation}
Then, we can write
\begin{equation}
    \mathcal{A} - \mathcal{A}_{0}  = -\frac{1}{2}\int_{0}^{\infty}\dd \tau \, \int_{-\infty}^{0}\dd \tau \,' \frac{g(\tau, \tau')}{(\Delta \tau - \ii \epsilon)^2} - \frac{1}{2}\int_{-\infty}^{0}\dd \tau \, \int_{0}^{\infty}\dd \tau \,' \frac{g(\tau, \tau')}{(\Delta \tau - \ii \epsilon)^2}.
\end{equation}
This can be further simplified by using $g(\tau, \tau') = g(\tau', \tau)^{*}$. Indeed,
\begin{equation}
    \mathcal{A} - \mathcal{A}_{0}  = -\frac{1}{2}\int_{0}^{\infty}\dd \tau \, \int_{-\infty}^{0}\dd \tau \,'\Biggl[\frac{g(\tau, \tau')}{(\Delta \tau - \ii \epsilon)^2} +  \frac{g(\tau, \tau')^{*}}{(\Delta \tau + \ii \epsilon)^2}\Biggr]
\end{equation}
Now, performing the change of variables $(\tau, \tau') \to (u, s)$, with $u = \tau - \tau'$ and $s = \tau$, we can write
\begin{equation}
    \mathcal{A} - \mathcal{A}_{0} = -\frac{1}{2}\int_{0}^{\infty}\dd s \int_{s}^{\infty}\dd u \, \chi(s)\chi(s-u)\Biggl[\frac{e^{\ii \Omega u}}{(u + \ii \epsilon)^2} + \frac{e^{-\ii \Omega u}}{(u - \ii \epsilon)^2}\Biggr].
\end{equation}
Changing the order of integration,
\begin{equation}
\mathcal{A} - \mathcal{A}_{0} = -\frac{1}{2}\int_{0}^{\infty}\dd u \, \int_{0}^{u}\dd s \chi(s)\chi(s-u)\Biggl[\frac{e^{\ii \Omega u}}{(u + \ii \epsilon)^2} + \frac{e^{-\ii \Omega u}}{(u - \ii \epsilon)^2}\Biggr].
\end{equation}
For the sake of organization, we define the auxiliary function
\begin{equation}
    \varphi(u) = \int_{0}^{u}\dd s \chi(s)\chi(s - u).
\label{eq:defvarphi}
\end{equation}
This way, we can compactly write
\begin{equation}
    \mathcal{A} - \mathcal{A}_{0} = -\frac{1}{2}\int_{0}^{\infty}\dd u \, \varphi(u) \Biggl[\frac{e^{\ii \Omega u}}{(u + \ii \epsilon)^2} + \frac{e^{-\ii \Omega u}}{(u - \ii \epsilon)^2}\Biggr].
    \label{eq:AminusA0compact}
\end{equation}
To proceed with the calculation, one might first expect that identity~\eqref{eq:generalizedSP2} could be applied once again, now with $n=2$. However, this identity is valid only for test functions defined on the whole real line. If one attempts to extend the integrand in Eq.~\eqref{eq:AminusA0compact} by multiplying it by $\Theta(u)$, the resulting function is not an admissible test function, since it is not differentiable at $u=0$. A more careful treatment is therefore required. Let $f(x)$ be any sufficiently smooth function with compact support defined on $[0, \infty)$, and consider
\begin{equation}
    I_{\epsilon} = \int_{-\infty}^{\infty}\dd x \, \frac{\Theta(x)f(x)}{x \pm \ii \epsilon} = \int_{0}^{\infty}\dd x \, \frac{f(x)}{x \pm \ii \epsilon}. 
\end{equation}
We would like to evaluate the limit $\lim_{\epsilon \to 0^{+}}I_{\epsilon}$.
Integrating by parts,
\begin{equation}
    I_{\epsilon} = -f(0)\log(\pm \ii \epsilon) - \int_{0}^{\infty}\dd x \, f'(x)\log(x \pm \ii \epsilon).
\end{equation}
Using
\begin{equation}
    \log z = \log |z| + \ii \operatorname{arg}z,
\end{equation}
we have
\begin{equation}
    I_{\epsilon} = -f(0)\log(\epsilon) \mp \frac{\ii \pi}{2}f(0) - \int_{0}^{\infty}\dd x \, f'(x)\left(\log(\sqrt{x^2 + \epsilon^2}) + \ii \operatorname{arg}(x \pm \ii \epsilon)\right).
\label{eq:Iepsilonintermediary}
\end{equation}
On the other hand, for every $\delta > 0$, we can write
\begin{equation}
\int_{\delta}^{\infty}\dd x \,\frac{f(x)}{x} = -\log(\delta)f(\delta) - \int_{\delta}^{\infty} \dd x \, \log x f'(x).
\label{eq:integralforcomparison}
\end{equation}
In the half-line, the $\operatorname{FP}$ distribution can be defined as
\begin{equation}
    \operatorname{FP} \int_{0}^{\infty}\dd x \, \frac{f(x)}{x} \equiv \lim_{\delta \to 0^{+}}\left[ \int_{\delta}^{\infty} \dd x \, \frac{f(x)}{x} + f(0)\log \delta\right].
    \label{eq:FP_particular_def}
\end{equation}
Hence,
\begin{equation}
\operatorname{FP}\int_{0}^{\infty}\dd x \, \frac{f(x)}{x} =  - \int_{0}^{\infty} \dd x \, f'(x) \log x.
\end{equation}
Furthermore, since $x > 0$, we have $\operatorname{arg} x = 0$, so that
\begin{equation}
    \lim_{\epsilon \to 0^{+}}\left(\log(\sqrt{x^2 + \epsilon^2}) + \ii \operatorname{arg}(x \pm \ii \epsilon)\right) =  \log x.
\end{equation}
Therefore,
\begin{equation}
    \lim_{\epsilon \to 0^{+}}\int_{0}^{\infty}\dd x \, f'(x)\left(\log(\sqrt{x^2 + \epsilon^2}) + \ii \operatorname{arg}(x \pm \ii \epsilon)\right) =  \int_{0}^{\infty}\dd x \, f'(x) \log x = -\operatorname{FP}\int_{0}^{\infty}\dd x \, \frac{f(x)}{x}.
\end{equation}
It then follows from Eq.~\eqref{eq:Iepsilonintermediary} that $I_{\epsilon}$ contains a logarithmically divergent contribution proportional to $\log(\epsilon)$, which cannot be discarded at this stage. We must therefore keep track of the divergence explicitly and verify whether it cancels in the final expression. Thus, we write
\begin{equation}
    \lim_{\epsilon \to 0^{+}}(I_{\epsilon} + f(0)\log(\epsilon)) = \mp\ii \pi \frac{f(0)}{2} + \operatorname{FP}\int_{0}^{\infty}\dd x \,\frac{f(x)}{x},
    \label{eq:Harveydent}
\end{equation}
which is the {\it Sokhotski–Plemelj theorem for the half-line}. Next, we need to evaluate
\begin{equation}
    \lim_{\epsilon \to 0^{+}}\int_{0}^{\infty}\dd x \,\frac{f(x)}{(x \pm \ii \epsilon)^2} \equiv \lim_{\epsilon \to 0^{+}} J_{\epsilon},
\end{equation}
which are precisely the integrals that appear in Eq.~\eqref{eq:AminusA0compact}. Once more, we start using integration by parts,
\begin{align}
   J_{\epsilon} & = -\frac{f(x)}{x\pm \ii \epsilon} \Bigg|^{\infty}_{0} + \int_{0}^{\infty}\dd x \, \frac{f'(x)}{(x \pm \ii \epsilon)} \nonumber \\
   & = \mp \ii \frac{f(0)}{\epsilon} + \int_{0}^{\infty}\dd x \, \frac{f'(x)}{(x \pm \ii \epsilon)}.
   \label{eq:Jepsilon}
\end{align}
Now, for each $\delta > 0$,
\begin{equation}
    \int_{\delta}^{\infty}\dd x \,\frac{f'(x)}{x} =-\frac{f(\delta)}{\delta} + \int_{\delta}^{\infty}\dd x \, \frac{f(x)}{x^2}.
\end{equation}
Using the Taylor expansion of $f(\delta)$ around $\delta = 0$, we have
\begin{equation}
    \frac{f(\delta)}{\delta} = \frac{f(0)}{\delta} + f'(0) + \mathcal{O}(\delta).
\end{equation}
Thus, in the limit $\delta \to 0$, the only non-vanishing and non-divergent term is $f'(0)$. Hence, we can conclude that
\begin{equation}
    \operatorname{FP} \int_{0}^{\infty}\dd x \, \frac{f'(x)}{x} = -f'(0) + \operatorname{FP}\int_{0}^{\infty}\dd x \, \frac{f(x)}{x^2}.
\end{equation}
Therefore, by combining Eqs.\eqref{eq:Harveydent} and \eqref{eq:Jepsilon} under the replacement $f \to f'$, we obtain
\begin{equation}
    \lim_{\epsilon \to 0^{+}} \Biggl( J_{\epsilon} + f'(0)\log \epsilon \pm \ii \frac{f(0)}{\epsilon}\Biggr) = \mp \ii \pi\frac{f'(0)}{2} - f'(0) + \operatorname{FP}\int_{0}^{\infty}\dd x \,\frac{f(x)}{x^2}.
    \label{eq:LeonSK}
\end{equation}
In distributional terms,
\begin{equation}
\lim_{\epsilon \to 0^{+}}\Biggl[\frac{\Theta(x)}{(x \mp\ii \epsilon)^2} - \log(\epsilon)\delta_{1/2}'(x) \pm2\ii\frac{\delta(x)\Theta(x)}{\epsilon}\Biggr] = \operatorname{FP}\left(\frac{\Theta(x)}{x^2} \right) + \delta_{1/2}'(x) \mp  \frac{\ii \pi}{2} \delta_{1/2}'(x),
\label{eq:GeneralizedrenromalizedFlamengo}
\end{equation}
where we have introduced the notation $\delta'_{1/2}(x)$ to emphasize that these distributions are normalized so that they do not acquire an additional factor of $1/2$ when integrated against a test function over the half-line. Concretely,
\begin{equation}
    \int_{0}^{\infty}\dd x \,  \delta_{1/2}'(x)f(x) = -f'(0).
\end{equation}
We can now apply Eq.~\eqref{eq:LeonSK} to solve Eq.~\eqref{eq:AminusA0compact}. For the sake of notational simplicity, consider
\begin{equation}
    f_{\pm}(u) = e^{\pm \ii \Omega u} \varphi(u).
\end{equation}
In this case, we can rewrite
\begin{equation}
    \mathcal{A} - \mathcal{A}_{0} = -\frac{1}{2}\int_{0}^{\infty}\dd u \, \frac{f_{+}(u)}{(u + \ii \epsilon)^2} -\frac{1}{2}\int_{0}^{\infty}\dd u \, \frac{f_{-}(u)}{(u - \ii \epsilon)^2}.
\end{equation}
Applying Eq.~\eqref{eq:LeonSK}, we can see that the complex-valued divergent terms cancel out. The remaining divergent term reads
\begin{equation}
    -\frac{1}{2}\log \epsilon (f_{+}'(0) + f_{-}'(0)) = -\chi(0)^{2}\log \epsilon
\end{equation}
Therefore,
\begin{align}
     \lim_{\epsilon \to 0^{+}}(\mathcal{A} -\mathcal{A}_{0} -\chi^{2}(0)\log \epsilon)& = \frac{1}{2}(f_{+}'(0) + f_{-}'(0)) -\frac{1}{2}\operatorname{FP}\int_{0}^{\infty}\dd u \, \frac{f_{-}(u) + f_{+}(u)}{u^2} \nonumber \\  &= \chi(0)^2 - \operatorname{FP}\int_{0}^{\infty}\dd u \, \frac{\varphi(u)\cos(\Omega u)}{u^2}.
\end{align}
To handle the $\operatorname{FP}$ integral, we can proceed as follows
\begin{align}
    & \operatorname{FP}\int_{0}^{\infty}\dd u \, \frac{\varphi(u)\cos(\Omega u)}{u^2} = \operatorname{FP}\int_{0}^{\infty}\dd u \, \frac{\cos(\Omega u)(\varphi(u) - u \varphi'(0) + u \varphi'(0))}{u^2} \nonumber \\ & = \int_{0}^{\infty}\dd u \, \frac{\cos(\Omega u)(\varphi(u) - u \chi(0)^2)}{u^2} + \chi(0)^2 \operatorname{FP}\int_{0}^{\infty}\dd u \, \frac{\cos(\Omega u)}{u},
\end{align}
as the first integral is regular. Therefore,
\begin{align}
     \lim_{\epsilon \to 0^{+}}(\mathcal{A} -\mathcal{A}_{0} -\chi^{2}(0)\log \epsilon) &  = -\int_{0}^{\infty}\dd u \, \frac{\cos(\Omega u)}{u^2}(\varphi(u) - u \chi(0)^2)  \nonumber \\ & -\chi(0)^2 \operatorname{FP}\int_{0}^{\infty}\dd u \, \frac{\cos(\Omega u)}{u} + \chi(0)^2.
     \label{eq:Afinalform}
\end{align}
Next, let us focus on simplifying Eq.~\eqref{eq:term_mathcal_B_expanded}. We start by writting
\begin{equation}
    \mathcal{B} = -\Biggl(I_{0} + \frac{I_{\tc{rf}}}{2} + \frac{I_{\tc{fr}}}{2}\Biggr),
    \label{eq:Bdefminussingconvenience}
\end{equation}
where
\begin{equation}
    I_{0} = \int_{-\infty}^{\infty}\int_{-\infty}^{\infty} \dd \tau \, \dd \tau \,'g(\tau, \tau')\frac{\partial_{\tau}f_{0}(\tau,\tau')-\partial_{\tau'}f_{0}(\tau,\tau')}{\Delta\tau-\ii\epsilon},
\end{equation}
\begin{equation}
    I_{\tc{rf}} = \int_{-\infty}^{\infty}\int_{-\infty}^{\infty} \dd \tau \, \dd \tau \,'g(\tau, \tau')\frac{\partial_{\tau}f_{\tc{rf}}(\tau,\tau')-\partial_{\tau'}f_{\tc{rf}}(\tau,\tau')}{\Delta\tau-\ii\epsilon},
\end{equation}
and
\begin{equation}
    I_{\tc{fr}} = \int_{-\infty}^{\infty}\int_{-\infty}^{\infty} \dd \tau \, \dd \tau \,'g(\tau, \tau')\frac{\partial_{\tau}f_{\tc{fr}}(\tau,\tau')-\partial_{\tau'}f_{\tc{fr}}(\tau,\tau')}{\Delta\tau-\ii\epsilon}.
\end{equation}
For the first integral, we have
\begin{align}
    I_{0} & =  \int_{-\infty}^{\infty}\int_{-\infty}^{\infty} \dd \tau \, \dd \tau \,' \frac{g(\tau, \tau')}{\Delta \tau - \ii \epsilon}[\Theta(-\tau')\Theta(\tau' + 2a)(\delta(\tau + 2a) - \delta(\tau)) + \delta(\tau)\Theta(\tau') \nonumber \\ & -\Theta(-\tau)\Theta(\tau + 2a)(\delta(\tau' + 2a) -\delta(\tau')) - \delta(\tau')\Theta(\tau)] \nonumber \\
    & = \int_{-\infty}^{0}\dd \tau \,' \frac{g(0, \tau')}{\tau' + \ii \epsilon} - \int_{0}^{\infty}\dd \tau \,' \frac{g(0, \tau')}{\tau' + \ii \epsilon} + \int_{-\infty}^{0}\dd \tau \, \frac{g(\tau, 0)}{\tau - \ii \epsilon} - \int_{0}^{\infty}\dd \tau \, \frac{g(\tau, 0)}{\tau -\ii \epsilon}
\end{align}
where, once more, we use the fact that $\chi(\tau) = 0$ for $\tau \leq -2a$ to extend the lower limit of the integrals from $-2a$ to $-\infty$. Next, using the definition of $g(\tau, \tau')$ and relabelling dummy variables, the integral $I_{0}$ can be cast into
\begin{align}
    I_{0}&  = \chi(0) \Biggl[\int_{-\infty}^{0} \dd u \, - \int_{0}^{\infty}\dd u \,\Biggr]\chi(u)\Biggl(\frac{e^{-\ii \Omega u}}{u - \ii \epsilon} + \frac{e^{\ii \Omega u}}{u + \ii \epsilon} \Biggr) \nonumber \\ & = -\chi(0)\int_{0}^{\infty}\dd u \, (\chi(u) + \chi(-u)) \Biggl(\frac{e^{-\ii \Omega u}}{u - \ii \epsilon} + \frac{e^{\ii \Omega u}}{u + \ii \epsilon} \Biggr).
\end{align}
As for the integral $I_{\tc{rf}}$, we have
\begin{align}
    I_{\tc{rf}} & = \int_{-\infty}^{\infty}\int_{-\infty}^{\infty} \dd \tau \, \dd \tau \,' \frac{g(\tau, \tau')}{\Delta \tau - \ii \epsilon}\left[\Theta(\tau')(\delta(\tau + 2a) - \delta(\tau)) - \delta(\tau')\Theta(-\tau)\Theta(\tau + 2a)\right] \nonumber \\ & =\int_{0}^{\infty}\dd \tau \,' \frac{g(0, \tau')}{\tau' + \ii \epsilon} - \int_{-\infty}^{0}\dd \tau \, \frac{g(\tau, 0)}{\tau - \ii \epsilon} \nonumber \\
    & = \chi(0)\Biggr[\int_{0}^{\infty}\dd u \, \frac{\chi(u)e^{\ii \Omega u}}{u + \ii \epsilon} - \int_{-\infty}^{0} \dd u \, \frac{\chi(u)e^{-\ii \Omega u}}{u - \ii \epsilon} \Biggl] \nonumber \\
    & = \chi(0)\int_{0}^{\infty}\dd u \, \frac{e^{\ii \Omega u}(\chi(u) + \chi(-u))}{u + \ii \epsilon}.
    \label{eq:I_rf_simplified}
\end{align}
Finally, the term $I_{\tc{fr}}$ can be cast into
\begin{align}
    I_{\tc{fr}} & = \int_{-\infty}^{\infty}\int_{-\infty}^{\infty} \dd \tau \, \dd \tau \,' \frac{g(\tau, \tau')}{\Delta \tau - \ii \epsilon}[\delta(\tau)\Theta(-\tau')\Theta(\tau' + 2a) - \Theta(\tau)(\delta(\tau' + 2a) - \delta(\tau'))] \nonumber \\
    & = -\int_{-\infty}^{0}\dd \tau \,' \frac{g(0, \tau')}{\tau' + \ii \epsilon} + \int_{0}^{\infty}\dd \tau \, \frac{g(\tau, 0)}{\tau - \ii \epsilon} \nonumber \\ 
    & = \chi(0)\Biggr[\int_{0}^{\infty}\dd u \, \frac{\chi(u)e^{-\ii \Omega u}}{u - \ii \epsilon} - \int_{-\infty}^{0} \dd u \, \frac{\chi(u)e^{\ii \Omega u}}{u + \ii \epsilon} \Biggl] \nonumber \\
    & = \chi(0)\int_{0}^{\infty}\dd u \, \frac{e^{-\ii \Omega u}(\chi(u) + \chi(-u))}{u - \ii \epsilon}.
    \label{eq:I_fr_simplified}
\end{align}
Combining Eqs.\eqref{eq:I_rf_simplified} and \eqref{eq:I_fr_simplified}, we obtain
\begin{equation}
    \frac{1}{2}(I_{\tc{rf}} + I_{\tc{fr}}) =  \frac{\chi(0)}{2}\int_{0}^{\infty}\dd u \, (\chi(u) + \chi(-u)) \Biggl[\frac{e^{-\ii \Omega u}}{u - \ii \epsilon} + \frac{e^{\ii \Omega u}}{u + \ii \epsilon}\Biggr].
\end{equation}
Thus, from Eq.~\eqref{eq:Bdefminussingconvenience}, it follows that (with $\chi_{\tc{s}}(u) \equiv \chi(u) + \chi(-u)$)
\begin{equation}
    \mathcal{B} = \frac{\chi(0)}{2}\int_{0}^{\infty}\dd u \, \chi_{\tc{s}}(u) \Biggl(\frac{e^{-\ii \Omega u}}{u - \ii \epsilon} + \frac{e^{\ii \Omega u}}{u + \ii \epsilon} \Biggr).
\end{equation}
Using Eq.~\eqref{eq:Harveydent},
\begin{align}
    \lim_{\epsilon \to 0^{+}} (\mathcal{B} + \chi(0)\chi_{\tc{s}}(0)\log \epsilon) &  = \frac{\ii \pi}{4}\chi_{\tc{s}}(0)\chi(0) -\frac{\ii \pi}{4}\chi_{\tc{s}}(0)\chi(0) + \frac{\chi(0)}{2}\chi(0)\operatorname{FP}\int_{0}^{\infty}\dd u \, \frac{\chi_{\tc{s}}(u)}{u}(e^{-\ii \Omega u} + e^{\ii \Omega u}) \nonumber \\
     & = \chi(0) \operatorname{FP}\int_{0}^{\infty}\dd u \, \frac{\cos(\Omega u)\chi_{\tc{s}}(u)}{u}.
\end{align}
Let us now evaluate the $\operatorname{FP}$ integral. The Taylor series for $\chi_{\tc{s}}(u)$ around $u = 0$ reads
\begin{equation}
    \chi_{\tc{s}}(u) = 2\chi(0) + \chi''(0)u^2 + \mathcal{O}(u^4).
\end{equation}
This means that the integral
\begin{equation}
    \int_{0}^{\infty}\dd u \, \frac{\cos(\Omega u) (\chi_{\tc{s}}(u) - 2\chi(0))}{u}
\end{equation}
is regular. Thus, it is possible to write
\begin{align}
    & \chi(0)\operatorname{FP}\int_{0}^{\infty}\dd u \, \frac{\cos(\Omega u)\chi_{\tc{s}}(u)}{u}  \nonumber \\ & = \chi(0)\int_{0}^{\infty}\dd u \, \frac{\cos(\Omega u)(\chi(u) + \chi(-u) - 2\chi(0))}{u} + 2\chi(0)^2\operatorname{FP} \int_{0}^{\infty}\dd u \, \frac{\cos(\Omega u)}{u}.
    \label{eq:FPBFinalform}
\end{align}
Hence,
\begin{align}
    \lim_{\epsilon \to 0^{+}}(\mathcal{B} + 2 \chi(0)^2 \log \epsilon) & = \chi(0)\int_{0}^{\infty}\dd u \, \frac{\cos(\Omega u)(\chi(u) + \chi(-u) - 2\chi(0))}{u} \nonumber \\ & + 2\chi(0)^2\operatorname{FP} \int_{0}^{\infty}\dd u \, \frac{\cos(\Omega u)}{u}.
    \label{eq:Bfinalform}
\end{align}
As for the final term, Eq.~\eqref{eq:term_mathcal_C_expanded}, we have
\begin{align}
    \mathcal{C} & = \frac{1}{2}\int_{-\infty}^{\infty} \dd \tau \,  \int_{-\infty}^{\infty} \dd \tau \,' g(\tau, \tau') \Biggl[ L_{0}(\Delta \tau) [2\delta(\tau)\delta(\tau') \nonumber \\ &  - \delta(\tau + 2a)\delta(\tau') - \delta(\tau' + 2a)\delta(\tau) + \delta(\tau + 2a)\delta(\tau' + 2a)] \nonumber \\ & +  L_{1}(\Delta \tau) [\delta(\tau')(\delta(\tau + 2a) - \delta(\tau))] \nonumber \\ & +  L_{1}(\Delta \tau) [\delta(\tau)(\delta(\tau' + 2a) - \delta(\tau'))]\Biggr] \nonumber \\ 
    &=  \frac{g(0, 0)}{2} [2L_{0}(0) -L_{1}(0) - L_{1}(0)] = \frac{\chi(0)^2}{2}(2 \log(\Lambda^2 \epsilon^2) - 2 \log(\Lambda \epsilon)) \nonumber \\
    & = \frac{\chi(0)^2}{2}(4 \log(\Lambda \epsilon) -2 \log(\Lambda \epsilon)) = \chi(0)^2\log(\Lambda \epsilon) = \chi(0)^2\log \epsilon + \chi(0)^2 \log (\Lambda).
    \label{eq:Cfinalform}
\end{align}
To show that the response function obtained via the method presented here matches the one obtained by Ref.~\cite{Louko2014}, let us define
\begin{equation}
    \Delta \mathcal{F}^{(\pi)}_{\tc{fw}} \equiv \mathcal{F}^{(\pi)}_{\tc{fw}} - \mathcal{F}^{(\pi)}_{0} =  -\frac{1}{2\pi}\lim_{\epsilon \to 0^{+}}(\mathcal{A} -\mathcal{A}_{0} + \mathcal{B} + \mathcal{C}).
\end{equation}
The divergent $\log\epsilon$ terms in Eqs.~\eqref{eq:Afinalform}, \eqref{eq:Bfinalform}, and \eqref{eq:Cfinalform} cancel exactly. We are therefore left with
\begin{align}
    \Delta\mathcal{F}^{(\pi)}_{\tc{fw}} & =  -
    \frac{\chi(0)^2}{2\pi} - \frac{\chi(0)^{2}\log(\Lambda)}{2\pi} +  \frac{1}{2\pi}\int_{0}^{\infty}\dd u \, \frac{\cos(\Omega u)(\varphi(u) - u \chi(0)^2)}{u^2} +\frac{\chi(0)^2}{2\pi} \operatorname{FP}\int_{0}^{\infty}\dd u \, \frac{\cos(\Omega u)}{u} \nonumber \\ 
    & -\frac{\chi(0)}{2\pi}\int_{0}^{\infty}\dd u \, \frac{\cos(\Omega u)(\chi(u) + \chi(-u) - 2\chi(0))}{u} -\frac{\chi(0)^2}{\pi}\operatorname{FP} \int_{0}^{\infty}\dd u \, \frac{\cos(\Omega u)}{u} \nonumber \\
    & = \frac{\chi(0)^2}{2\pi}\log(e^{-1}/\Lambda) + \frac{1}{2\pi}\int_{0}^{\infty}\dd u \, \cos(\Omega u)\left[\frac{\chi(0)(\chi(0) - \chi(u) - \chi(-u))}{u}  + \frac{\varphi(u)}{u^2}\right]\nonumber \\
    & -\frac{\chi(0)^2}{2\pi}\operatorname{FP} \int_{0}^{\infty}\dd u \, \frac{\cos(\Omega u)}{u}
    \label{eq:final_result_form_1}
\end{align}
To evaluate the $\operatorname{FP}$ of the cosine integral, consider
\begin{equation}
    \operatorname{FP} \int_{0}^{\infty}\dd u \, \frac{\cos(\Omega u)}{u} = \operatorname{FP}\int_{0}^{1}\dd u \, \frac{\cos(\Omega u)}{u} + \int_{1}^{\infty} \dd u \, \frac{\cos(\Omega u)}{u},
    \label{eq:anotherdayanotherPV}
\end{equation}
where, in the last integral, we have dropped the $\operatorname{FP}$ notation since the expression is already convergent in the ordinary sense. We also recall that, for the singular contribution $u=0$, $\operatorname{FP}$ is understood in the sense of Eq.~\eqref{eq:FP_particular_def}. We may then use the identity
\begin{equation}
    \int_{x}^{\infty}\dd t \,\frac{\cos t}{t} = -\gamma - \log x - \int_{0}^{x}\dd t \,\frac{\cos t - 1}{t},
    \label{eq:of_course}
\end{equation}
where $\gamma$ is the Euler–Mascheroni constant. Taking $x = 1$ and assuming $\Omega \neq 0$, we have
\begin{align}
 \int_{1}^{\infty}\dd u \, \frac{\cos(\Omega u)}{u} & = \int_{|\Omega|}^{\infty}\dd t \, \frac{\cos t}{t} = -\gamma - \log(|\Omega|) - \int_{0}^{|\Omega|}\dd t \, \frac{\cos t - 1}{t} \nonumber \\
 & =  -\gamma - \log(|\Omega|) - \int_{0}^{1}\dd u \, \frac{\cos (\Omega u) - 1}{u}. 
 \label{eq:outofideas}
\end{align}
Thus,
\begin{equation}
    \operatorname{FP} \int_{0}^{\infty}\dd u \, \frac{\cos(\Omega u)}{u} = -\gamma - \log(|\Omega|) + \operatorname{FP}\int_{0}^{1}\dd u \, \frac{\cos(\Omega u)}{u} - \int_{0}^{1}\dd u \, \frac{\cos(\Omega u) - 1}{u}. 
\end{equation}
We now claim that
\begin{equation}
    \operatorname{FP}\int_{0}^{1}\dd u \, \frac{\cos(\Omega u)}{u} = \int_{0}^{1}\dd u \, \frac{\cos(\Omega u) - 1}{u}.
    \label{eq:Pv01cosoveru}
\end{equation}
Indeed, for every $\delta > 0$,
\begin{align}
    \operatorname{FP}\int_{0}^{1}\dd u \, \frac{\cos(\Omega u)}{u} & =  \operatorname{PV}\left(\int_{0}^{\delta}\dd u \, \frac{\cos(\Omega u)}{u} + \int_{\delta}^{1}\dd u \, \frac{\cos(\Omega u)}{u} \right) \nonumber \\
    & =\operatorname{FP}\left( \int_{0}^{\delta}\dd u \, \frac{\cos(\Omega u)}{u} + \int_{\delta}^{1}\dd u \, \frac{\cos(\Omega u) - 1}{u} + \int_{\delta}^{1}\frac{\dd u \, }{u}\right) \nonumber \\
   & = \operatorname{FP}\left( \int_{0}^{\delta}\dd u \, \frac{\cos(\Omega u)}{u} + \int_{\delta}^{1}\dd u \, \frac{\cos(\Omega u) - 1}{u} - \log \delta\right).
\end{align}
From this, one concludes that Eq.~\eqref{eq:Pv01cosoveru} holds. Therefore,  
\begin{equation}
     \operatorname{FP} \int_{0}^{\infty}\dd u \, \frac{\cos(\Omega u)}{u} = -\gamma - \log(|\Omega|).    
\end{equation}
Finally, the result in Eq.~\eqref{eq:final_result_form_1} can be rewritten as
\begin{align}
    \Delta \mathcal{F}^{(\pi)}_{\tc{fw}} & = \frac{\chi(0)^2}{2\pi}\log(|\Omega|e^{\gamma - 1}/\Lambda) \nonumber \\ & +  \frac{1}{2\pi}\int_{0}^{\infty}\dd u \, \cos(\Omega u)\left[\frac{\chi(0)(\chi(0) - \chi(u) - \chi(-u))}{u}  + \frac{\varphi(u)}{u^2}\right],
    \label{eq:requiem}
\end{align}
which is precisely the one obtained by Ref.~\cite{Louko2014}.

\section{State-relative normal ordering for Gaussian zero-mean states}
\label{sec:normal_ordering_general}

Let $\hat{A}(\mf x)$ denote a generic operator that depends on the massless scalar field $\hat{\phi}(\mf x)$ defined on Minkowski spacetime, e.g., $\hat{A}=\hat{\phi}^2$ or $\hat{A}=\hat{\pi}^2$. Consider a generic state $\hat{\rho}$ of the field. Then, depending on the form of $\hat{A}$, the correlation function
\begin{equation}
    \operatorname{Tr}[\hat{A}(\mf x)\hat{A}(\mf x')\hat{\rho}]
    \label{eq:toberenormalized}
\end{equation}
may require regularization. To describe the relevant formal algebra, given a reference state $\hat{\omega}$ of the field, we introduce the state-relative subtraction
\begin{equation}
    \mathopen{:}\hat{A}(\mf x)\mathopen{:}_{\omega}
    \equiv
    \hat{A}(\mf x) - \operatorname{Tr}[\hat{A}(\mf x)\hat{\omega}]\,\openone.
    \label{eq:normal_ordering_modified}
\end{equation}

The calculations carried out in Appendix~\ref{sec:appendix_two_point} to evaluate the energy-density correlation function assumed the vacuum as the reference state, $\hat{\omega} = \proj{0}{0}$. In this case, the formal algebraic expansion of the vacuum-normal-ordered correlation in Eq.~\eqref{eq:toberenormalized} reads
\begin{align}
    \operatorname{Tr}[\mathopen{:}\hat{A}(\mf x)\mathopen{:}\mathopen{:}\hat{A}(\mf x')\mathopen{:}\hat{\rho}]
    &=
    \operatorname{Tr}[\hat{A}(\mf x)\hat{A}(\mf x')\hat{\rho}]
    - \bra{0}\hat{A}(\mf x)\ket{0}\operatorname{Tr}[\hat{A}(\mf x')\hat{\rho}]
    \nonumber \\
    &\quad
    - \bra{0}\hat{A}(\mf x')\ket{0}\operatorname{Tr}[\hat{A}(\mf x)\hat{\rho}]
    + \bra{0}\hat{A}(\mf x)\ket{0}\bra{0}\hat{A}(\mf x')\ket{0}.
    \label{eq:general_renorm_correlations}
\end{align}
Let us consider the particular case $\hat{A}=\hat{\phi}^2$, and assume that $\hat{\rho}$ is a zero-mean Gaussian state, so that its $n$-point functions are determined by Wick's theorem and all odd-point functions vanish. Using Wick's theorem, we can write
\begin{equation}
    \operatorname{Tr}[\hat{\phi}(\mf x)^2 \hat{\phi}(\mf x')^2 \hat{\rho}]
    =
    W_{\rho}(\mf x, \mf x)W_{\rho}(\mf x', \mf x')
    + 2\bigl(W_{\rho}(\mf x, \mf x')\bigr)^2,
\end{equation}
where $W_{\rho}(\mf x, \mf x')$ denotes the Wightman function of the field in the state $\hat{\rho}$. Then, using Eq.~\eqref{eq:general_renorm_correlations}, we obtain
\begin{align}
    \operatorname{Tr}[\mathopen{:}\hat{\phi}^2(\mf x)\mathopen{:}\mathopen{:}\hat{\phi}^2(\mf x')\mathopen{:}\hat{\rho}]
    &=
    W_{\rho}(\mf x, \mf x)W_{\rho}(\mf x', \mf x')
    + 2W_{\rho}(\mf x, \mf x')^2
    \nonumber \\
    &\quad
    - W_{0}(\mf x, \mf x)W_{\rho}(\mf x', \mf x')
    - W_{0}(\mf x', \mf x')W_{\rho}(\mf x, \mf x)
    + W_{0}(\mf x, \mf x)W_{0}(\mf x', \mf x').
    \label{eq:modelA2correlationquasifree}
\end{align}
This expression is formal as written: the coincident quantities are not asserted to exist separately. To interpret its decomposition, one must fix a joint prescription for the full collection of quantities---for example, a regulator applied before the decomposition---and use it consistently in every one- and two-point term. Defining the separate terms independently could obscure cancellations present in the complete correlation.

For a general field state $\hat{\rho}$, vacuum normal ordering does not by itself ensure that Eq.~\eqref{eq:modelA2correlationquasifree} defines a bona fide distribution, since the differences between the coincident two-point functions need not be smooth. If $\hat{\rho}$ is a Hadamard state~\cite{fullingHadamard, HadamardBible, fewsterNecessityHadamard}, the difference $W_{\rho}-W_{0}$ is smooth, and Eq.~\eqref{eq:modelA2correlationquasifree} can be reduced to a well-defined distribution. However, for the firewall state considered in the main text, the Hadamard assumption fails. Consequently, the preceding Hadamard argument no longer guarantees that the complete vacuum-normal-ordered correlation admits a canonical, regulator-independent extension. This does not, by itself, prove that no such extension exists: a single joint prescription could in principle produce cancellations between the connected-correlation and the product of one-point functions.

Conditional on assigning all the relevant quantities a meaning through the same prescription, one may use the notation associated with $\hat{\omega}=\hat{\rho}$ as a formal state-relative subtraction. Algebraically, this subtraction represents the connected combination because
\begin{equation}
\mathopen{:}\hat{A}(\mf x)\mathclose{:}_{\rho}
=
\mathopen{:}\hat{A}(\mf x)\mathclose{:}
-
\operatorname{Tr}\!\left[
\mathopen{:}\hat{A}(\mf x)\mathclose{:}\hat{\rho}
\right]\openone ,
\end{equation}
and therefore
\begin{align}
\operatorname{Tr}\!\left[
\mathopen{:}\hat{A}(\mf x)\mathclose{:}_{\rho}
\mathopen{:}\hat{A}(\mf x')\mathclose{:}_{\rho}
\hat{\rho}
\right]
 =
\operatorname{Tr}\!\left[
\mathopen{:}\hat{A}(\mf x)\mathclose{:}
\mathopen{:}\hat{A}(\mf x')\mathclose{:}
\hat{\rho}
\right]
-
\operatorname{Tr}\!\left[
\mathopen{:}\hat{A}(\mf x)\mathclose{:}\hat{\rho}
\right]
\operatorname{Tr}\!\left[
\mathopen{:}\hat{A}(\mf x')\mathclose{:}\hat{\rho}
\right].
\end{align}
This notation is not asserted to define a bona fide, regulator-independent normal-ordered local operator for the sharp firewall state. If used to define the interaction Hamiltonian, it would instead produce a physically inequivalent detector model through the corresponding reference-state-dependent detector-only term. For $\hat{A}=\hat{\phi}^{2}$ and a Gaussian state with vanishing one-point function, Wick factorization formally gives
\begin{equation}
    \operatorname{Tr}[\mathopen{:}\hat{\phi}^2(\mf x)\mathopen{:}_{\rho}\mathopen{:}\hat{\phi}^2(\mf x')\mathopen{:}_{\rho}\hat{\rho}]
    =
    2 W_{\rho}(\mf x, \mf x')^2.
    \label{eq:newrenorm}
\end{equation}

The same formal algebra applies in the case $\hat{A} = \hat{\pi}^2 = (\partial_{\tau}\hat{\phi})^2$, which is the relevant operator in the model considered in Sec.~\ref{sec:andthenitdiverges}. Concretely, in Eqs.~\eqref{eq:modelA2correlationquasifree}--\eqref{eq:newrenorm}, the Wightman functions $W_{\rho}$ and $W_{0}$ should be replaced by
\begin{equation}
    \operatorname{Tr}[\partial_{\tau}\hat{\phi}(\mf x)\partial_{\tau'}\hat{\phi}(\mf x')\hat{\rho}]
    =
    \partial_{\tau}\partial_{\tau'}W_{\rho}(\mf x, \mf x'),
\end{equation}
and
\begin{equation}
    \operatorname{Tr}[\partial_{\tau}\hat{\phi}(\mf x)\partial_{\tau'}\hat{\phi}(\mf x')\proj{0}{0}]
    =
    \partial_{\tau}\partial_{\tau'}W_{0}(\mf x, \mf x').
\end{equation}
For a non-Hadamard state such as the firewall state, vacuum normal ordering does not by itself guarantee that the complete correlation defines a bona fide distribution. To represent the connected combination formally, suppose that all terms are assigned a meaning by the same prescription and set $\hat{\omega}=\hat{\rho}$. For a Gaussian state with vanishing one-point function, Wick factorization then formally gives
\begin{equation}
    \operatorname{Tr}\!\left[
    \mathopen{:}\hat{\pi}^2(\mf x)\mathclose{:}_{\rho}
    \mathopen{:}\hat{\pi}^2(\mf x')\mathclose{:}_{\rho}
    \hat{\rho}
    \right]
    =
    2\bigl(\partial_{\tau}\partial_{\tau'}W_{\rho}(\mf x,\mf x')\bigr)^2.
\end{equation}
For $\hat{\rho}=\hat{\rho}_{\tc{fw}}$, this is the formal state-relative representation of the connected correlation used in Eq.~\eqref{eq:response_fw_pi2}. Its use there does not amount to replacing the Minkowski-normal-ordered interaction Hamiltonian by a firewall-normal-ordered one.

\twocolumngrid

\bibliography{references}

\end{document}